\documentclass[journal]{IEEEtran}

\usepackage{filecontents}
\usepackage[noadjust]{cite}

\ifCLASSINFOpdf
\else
\fi
\usepackage{amsmath}
\usepackage{amssymb}
\usepackage{mathtools}
\usepackage{enumitem}
\usepackage{adjustbox}

\usepackage{array}
\usepackage{makecell}
\usepackage{multirow}
\newcolumntype{M}[1]{>{\centering\arraybackslash}m{#1}}

\usepackage{amsthm} 
\newtheorem{theorem}{Theorem} 

\usepackage[noadjust]{cite} 

\usepackage{graphicx}
\usepackage{array}
\usepackage{bm}
\usepackage{circuitikz}
\usepackage{tikz}
\usepackage{siunitx} 
\usepackage[hypcap]{caption}
\usetikzlibrary{arrows}
\usepackage{algorithm}
\usepackage{algorithmicx}
\usepackage{algpseudocode}
\usepackage{float}
\usepackage{subcaption}  

\usepackage[colorlinks=true, linkcolor=blue, citecolor=blue, urlcolor=blue]{hyperref}
\DeclareMathOperator*{\argmin}{\,min}

\newcommand{\RNum}[1]{\uppercase\expandafter{\romannumeral #1\relax}}

\begin{document}

\title{Compressed Sensing Based Residual Recovery Algorithms and Hardware for Modulo Sampling}

\author{
Shaik~Basheeruddin~Shah,~\IEEEmembership{Student~Member,~IEEE,}~Satish Mulleti,~\IEEEmembership{Member,~IEEE,}, Yonina~C.~Eldar,~\IEEEmembership{Fellow,~IEEE}
~
\thanks{Part of this work has been accepted for presentation at the IEEE International Conference on Acoustics, Speech, and Signal Processing (ICASSP) 2023.
S. B. Shah is with the Department of Electrical Engineering, Khalifa University, UAE.
S. Mulleti is with the Department of Electrical Engineering, Indian Institute of Technology (IIT) Bombay, India.
Y. C. Eldar is with the Weizmann Institute of Science, Rehovot, Israel. 
This research was supported by the Tom and Mary Beck Center for Renewable Energy as part of the Institute for Environmental Sustainability (IES) at the Weizmann Institute of Science, by the European Research Council (ERC) under the European Union’s Horizon 2020 research and innovation program (grant No. 101000967) and by the Israel Science Foundation (grant No. 536/22).}}
\maketitle

\begin{abstract}
Analog-to-Digital Converters (ADCs) are essential components in modern data acquisition systems. A key design challenge is accommodating high Dynamic Range (DR) input signals without clipping. Existing solutions such as oversampling, Automatic Gain Control (AGC), and compander-based methods have limitations in handling high DR signals. Recently, the Unlimited Sampling Framework (USF) has emerged as a promising alternative. It uses a non-linear modulo operator to map high-DR signals within the ADC’s range. 
Existing recovery algorithms, such as Higher-Order Differences (HODs), prediction-based, and Beyond Bandwidth Residual Recovery (B$^2$R$^2$) have shown potential but are either noise-sensitive, require high sampling rates, or are computationally intensive. To address these challenges, we propose LASSO-B$^2$R$^2$, a fast and robust recovery algorithm. Specifically, we demonstrate that the first-order difference of the residual (i.e., the difference between the folded and original samples) is sparse and derive an upper bound on its sparsity. This insight allows us to formulate the recovery as a sparse signal reconstruction problem using the Least Absolute Shrinkage and Selection Operator (LASSO).
Numerical simulations demonstrate that LASSO-B$^2$R$^2$ outperforms prior methods in terms of speed and robustness, although with a higher sampling rate at lower DR. To overcome this, we introduce the \textit{bits distribution} mechanism that allocates 1-bit from the total bit budget to identify modulo folding events. This reduces the recovery problem to a simple pseudo-inverse computation, significantly enhancing computational efficiency. Finally, we validate the superiority of our approach through numerical simulations and a hardware prototype that captures 1-bit folding information, demonstrating its practical feasibility.
\end{abstract}


\IEEEpeerreviewmaketitle

\section{Introduction}
Modern data acquisition systems relay on Analog-to-Digital Converters (ADCs) to convert continuous-time analog signals into discrete-time digital signals for processing.
ADCs typically operate by sampling the continuous-time signal at equally spaced intervals and then quantizing them to have finite precision \cite{Eldar}. 
The theory underlying this process is rooted in the Shannon-Nyquist sampling theorem, which states that a Band-Limited (BL) signal can be accurately reconstructed from its equally spaced samples if the sampling rate is at least twice the maximum frequency present in the signal \cite{Eldar}. This minimum rate is known as the Nyquist rate. 

A key consideration in an ADC design is ensuring that the Dynamic Range (DR) of the ADC is greater than that of the input analog signal. This avoids clipping of the signals. However, in practice, the DR of the input analog signal may surpass the ADC's DR. The resulting clipping results in information loss, and it is not possible to recover the original signal from the Nyquist samples.
This issue is a major concern in applications such as imaging \cite{HDR1, Method8, Method9}, robust vision systems \cite{HDR2}, multi-band communication \cite{Method2, Method10}, and seismic waveform analysis \cite{HDR3}. Therefore, to achieve optimal performance, the ADC must be designed to handle the full DR of the input signal effectively.

Numerous hardware preprocessing steps and algorithms have been proposed in the literature to address the DR constraints in ADCs \cite{Method0, Method1, Method3, Lit_4, Method4, Method5, Method6, Method7}.
Oversampling techniques for BL signals have been explored in \cite{Method0, Method1}, utilizing sample correlation to recover clipped data. For multi-band communication systems, prior knowledge of vacant bands is leveraged to reconstruct clipped samples \cite{Method3, Lit_4}. Additionally, preprocessing methods, such as Automatic Gain Control (AGC) circuits \cite{Method4, Method5}, and compander-based approaches \cite{Method6, Method7} have been investigated. While these methods offer distinct advantages, each has its limitations.
\begin{figure}[t!]
    \centering
    \includegraphics[height = 5.5cm, width = 8cm]{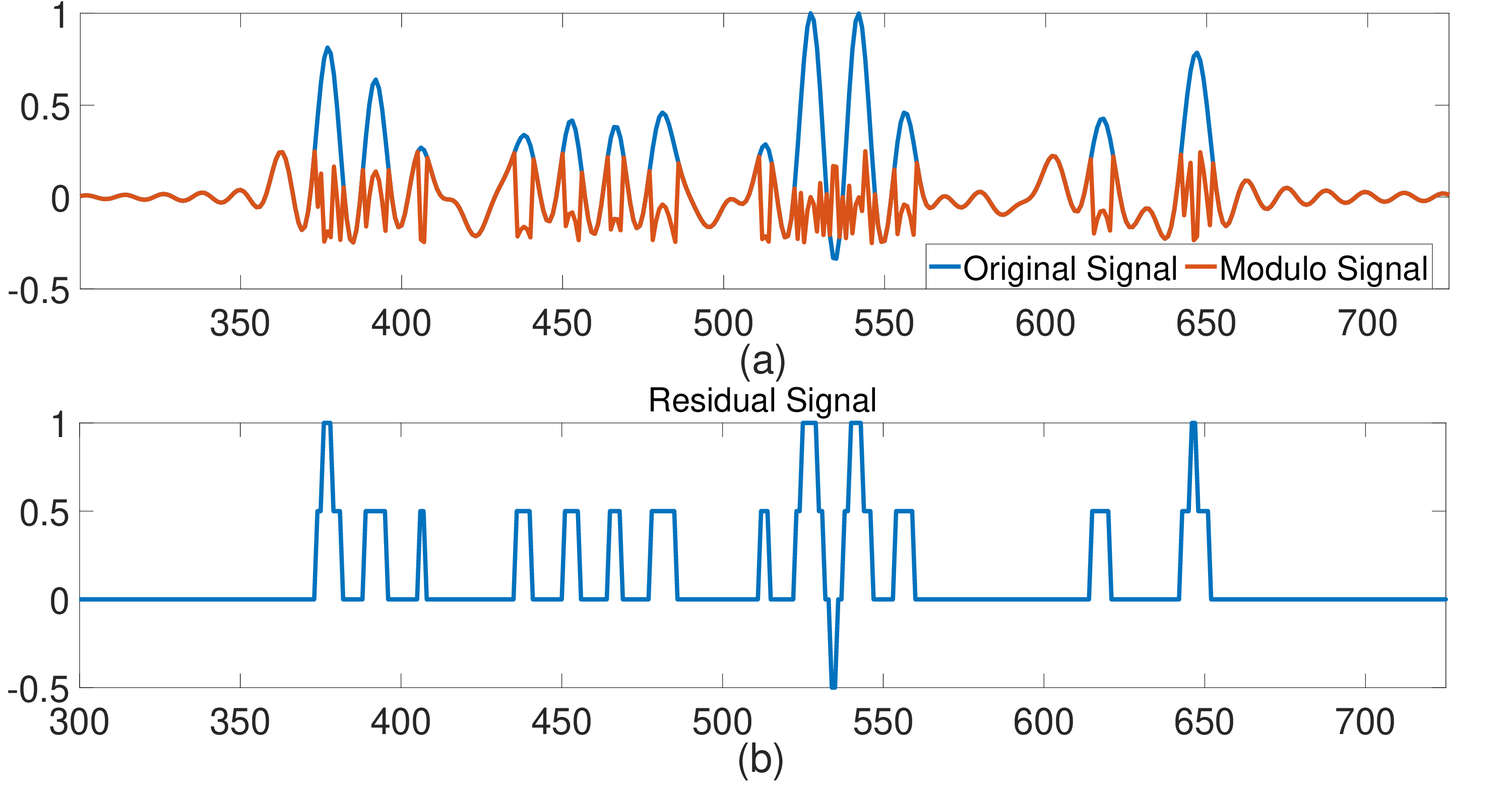}
    \caption{\small{(a) BL signal $f(t)$ and the corresponding modulo signal $f_\lambda(t)$ with $\lambda = 0.25$. (b) Residual signal, $z(t) = f_\lambda(t)-f(t)$. The samples between [300, 750] are displayed for better visibility.}}
    \label{Typical_Modulo_Signal}
\end{figure}

\begin{figure}[t!]
    \centering
    \includegraphics[height = 2cm, width = 8cm]{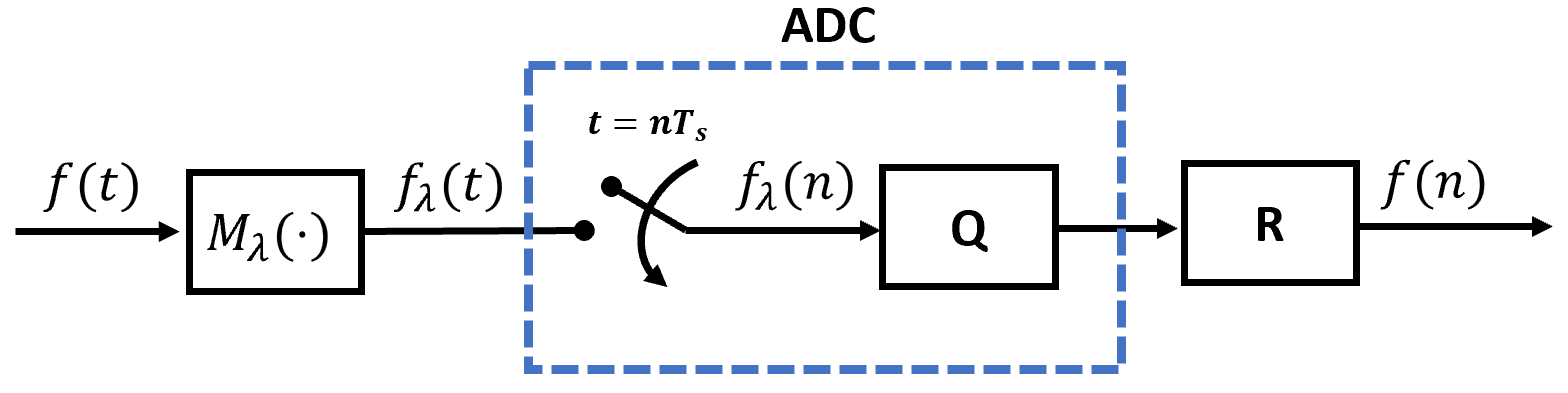}
    \caption{\small{Modulo-ADC block diagram.}}  \label{Modulo_Sampling_Block_Diagram}
\end{figure}

Improving upon the aforementioned methods, the use of a non-linear modulo operator for preprocessing high DR signals has gained significant attention \cite{Method8, Method9, Method11, Method12, US2, Cheb, B2R2, B2R21, Fernández, Modulo_MIMO, Fernández1, UNO}. This approach applies the modulo operation to the high-DR input signal, effectively reducing its DR to fit within the ADC's range. In the context of imaging, such high DR ADCs are called \textit{self-reset} ADCs \cite{Method8, Method9, Method11, Method12}. Among these modulo-based approaches, the recent Unlimited Sampling Framework (USF) has emerged as a particularly promising solution \cite{US2}.
USF begins by preprocessing the high DR input signal $f(t)$ using a modulo operation $\mathcal{M}_\lambda(.)$.
Fig. \ref{Typical_Modulo_Signal}(a) illustrates a typical high DR input signal, \( f(t) \), and its corresponding modulo signal, $f_\lambda(t)$. Note that the residue signal $z(t) = f_\lambda(t) - f(t)$ is a piecewise constant signal with its values in $2\lambda \mathbb{Z}$ (cf. Fig. \ref{Typical_Modulo_Signal}(b)).
After folding, the modulo signal is fed into the ADC, resulting in modulo samples. To reconstruct original samples from these modulo samples, a recovery algorithm, also called unfolding, $\mathcal{R}$, is integrated into the framework. 
The combined system of the ADC, modulo preprocessing, and recovery algorithm constitutes what is known as a modulo-ADC system, depicted in Fig. \ref{Modulo_Sampling_Block_Diagram}. This integrated framework enables effective sampling and reconstruction of high DR signals.

USF has been investigated across diverse signal models, including Finite-Rate-of-Innovation (FRI) signals \cite{Mulleti:2024}, 
sparse signals \cite{Musa:2018,Prasanna:2021,Shah:2021}, periodic BL signals \cite{Fourier-Prony}, graph signals \cite{Feng:2022}, wavelet representations \cite{Rudresh:2018}, 
shift-invariant spaces \cite{Kvich:2024, Yhonatan1} and more.
Hardware prototypes of modulo-ADCs are presented in \cite{Fourier-Prony, Mod_HW, Satish_HW, kvich2024practical}. 
Bhandari et al. \cite{US2} introduced the USF to recover BL unfolded samples by computing Higher-Order Differences (HODs) of the folded samples. However, this algorithm requires a high sampling rate, approximately $2\pi e$ times the Nyquist rate, where $e$ is Euler’s constant. Additionally, the reliance on HODs makes the method highly sensitive to noise.
In \cite{Fourier-Prony}, the authors proposed a spectral estimation method called Fourier-Prony, which uses the separation of the true signal's spectrum and the spectrum of the folded samples. The algorithm is shown to be useful for periodic-BL signals, and FRI signals \cite{AyushHW}.
In \cite{Cheb}, the authors proposed a prediction-based algorithm, demonstrating that perfect recovery of finite-energy BL signals is feasible when sampling above the Nyquist rate. This approach leverages the observation that for finite-energy BL signals, the amplitude becomes negligible beyond a certain time instant. Consequently, the folded samples and unfolded samples become identical beyond this point.
Building on this, the authors of \cite{B2R2, B2R21} introduced an improved and robust recovery algorithm called Beyond Bandwidth Residual Recovery (B$^2$R$^2$). This method reformulates the recovery of residual samples as a constrained optimization problem. Although B$^2$R$^2$ is noise robust compared to the existing methods, it is computationally expensive. Specifically, the algorithm recovers residual samples individually, solving a constrained optimization problem for each sample using an iterative procedure.
This motivates the need for a robust and computationally efficient algorithm that operates at lower sampling rates.

This paper aims to fill this gap by proposing fast and robust recovery algorithms that operate near the Nyquist rate.
The main contributions of this work are as follows:
\begin{itemize}
    \item We begin by demonstrating that the first-order difference of the residual samples, denoted as $\hat{z}(n)$, exhibits sparsity. This is theoretically justified by deriving an upper bound on the number of non-zero elements in $\hat{z}(n)$. Leveraging this observation, we formulate the recovery of residual samples as a sparse recovery problem. Specifically, we cast it as a Least Absolute Shrinkage and Selection Operator (LASSO) problem and solve it using the standard Iterative Soft-Thresholding Algorithm (ISTA). Once the residual samples are estimated, the true samples can be recovered from the given modulo samples. For BL signals, we incorporate the beyond-the-bandwidth frequency information of the modulo samples into the LASSO formulation, leading to the proposed LASSO-B$^2$R$^2$.
    \item Through numerical simulations, we demonstrate that LASSO-B$^2$R$^2$ outperforms existing methods in terms of speed and robustness. However, at lower $\lambda$ values, a slightly higher sampling rate is necessary. To address this limitation, we introduce the concept of \textit{bits distribution}. Specifically, we allocate 1-bit from the total available bit budget to detect the occurrence of modulo folding. This reduces the sparse recovery problem to a simple least squares problem with a closed-form solution.  Numerically, we demonstrate that the LASSO-B$^2$R$^2$ with 1-bit addresses the limitation of LASSO-B$^2$R$^2$. Additionally, we present a hardware prototype to obtain the 1-bit information in practice. Specifically, we show that one can use the existing hardware developed in \cite{Mod_HW} to extract the additional 1-bit information. Importantly, no additional circuitry is required, except for an OR gate.   
\end{itemize}
The LASSO-B$^2$R$^2$ with 1-bit, presented in this study, was utilized in \cite{Bernardo} to evaluate the performance of a modulo-ADC in comparison to a standard ADC. Specifically, \cite{Bernardo} analyzed the modulo-ADC within a dithered quantization framework, deriving a closed-form expression for the Mean-Square Error (MSE) and comparing it with that of the standard ADC. The analysis showed that an oversampling factor greater than 3 and a quantizer resolution exceeding 3 bits are sufficient for the modulo ADC to achieve better quantization noise suppression than a standard ADC.

The rest of the paper is organized as follows: Section \RNum{2} presents essential preliminaries necessary for the subsequent discussion and formulates the problem. Section \RNum{3} introduces the LASSO-B$^2$R$^2$ algorithm and provides a numerical comparison with existing methods. Section \RNum{4} explores the concept of \textit{bits distribution}, highlighting its advantages through numerical simulations and a hardware prototype demonstration.

Throughout the paper, we employ the following notations.
The space representing continuous-time BL signals with a bandwidth of $[-f_m, f_m]$ is denoted as $\mathcal{B}_{f_m}$.
The first-order difference operator of a signal $f(n)$ is $\hat{f}(n) = \Delta f(n)$. 
To ensure that $f(n)$ and $\hat{f}(n)$ have the same length during the computation of the first-order difference, we prepend $f(n)$ with a zero.
The symbols $\mathbb{R}$, $\mathbb{R}^+$, and $\mathbb{Z}$ denote the sets of real numbers, positive real numbers, and integers, respectively.
The symbol $[L]$ is the set $\{0, 1,\dots,L-1\}$.
The symbol $L^2(\mathbb{R})$ is the space of square-integrable functions on $\mathbb{R}$.
Bold letters are used to represent vectors and matrices.
The symbols $||\mathbf{z}||_1$, $||\mathbf{z}||_2$, and $||\mathbf{z}||_\infty$ represent the $l_1$-norm, $l_2$-norm, and $l_\infty$-norm of $\mathbf{z}$, respectively. The spectral norm of a matrix $\mathbf{V}$ is denoted as $||\mathbf{V}||_2$.
The pseudo-inverse of a matrix $\mathbf{A} \in \mathbb{R}^{m \times n}$, where $n\gg m$, is written as $\mathbf{A}^{\dagger}$, and defined as $\mathbf{A}^{\dagger} = (\mathbf{A}^T\mathbf{A})^{-1}\mathbf{A}^T$.

\section{Preliminaries and Problem Formulation}
\label{Preliminaries_and_Problem_Formulation}

\subsection{Preliminaries}
\label{Preliminaries}
Let $f(t)$ be a continuous time, finite-energy, BL, aperiodic signal, i.e., $f(t) \in L^2 (\mathbb{R})\cap \mathcal{B}_{\omega_m},\ t\in\mathbb{R}$.
Since modern signal processing systems operate in the digital domain, an ADC is employed to convert $f(t)$ into the digital domain.
Assume that the DR of $f(t)$, say $[-A, A]$, is greater than the DR of the ADC, say $[-\lambda, \lambda]$, i.e., $A>\lambda$. 
In such cases, conventional ADCs introduce signal clipping, leading to information loss.
To overcome this limitation, we employ the non-linear modulo operator, $\mathcal{M}_\lambda(\cdot)$, which is defined as
\begin{equation}
  \mathcal{M}_\lambda(x) = (x+\lambda)\ \text{mod}\ 2\lambda - \lambda,  
\end{equation}
where $x\in \mathbb{R}$ and the ADC's DR is $[-\lambda, \lambda]$ for some $\lambda>0$.  The modulo-folded signal, $f_\lambda(t) = \mathcal{M}_\lambda \left( f(t)\right)$, has the same DR as the ADC's. Specifically, we have that $|f_\lambda(t)| \leq \lambda$. 
Mathematically, the modulo signal is decomposed as 
\begin{equation}
    f_\lambda(t) = \mathcal{M}_\lambda(f(t)) = f(t) + z(t), \ t\in\mathbb{R},
    \label{Modulo_Equation}
\end{equation}
where $z(t)\in 2\lambda\mathbb{Z}$ represents the residual signal.
A typical residual signal is illustrated in Fig. \ref{Typical_Modulo_Signal}(c).
The folded signal $f_\lambda(t)$ is then sampled at regular uniform time intervals of $T_s$, resulting in the following modulo samples:
\begin{equation}
   f_\lambda(nT_s) = f_\lambda(n) = f(n) + z(n),\ n\in\mathbb{Z} .
   \label{Modulo_Equation_Discrete}
\end{equation}
In this context, the sampling rate, $f_s$, is expressed as
$f_s = \frac{1}{T_s} = \text{OF}\times 2f_m,$ where $\text{OF}>1$ denotes the oversampling factor and $w_m = 2f_m$ is the Nyquist rate.
The next step involves recovering the original (unfolded) samples \( f(n) \) from the modulo samples \( f_\lambda(n) \) using a recovery algorithm \( \mathcal{R} \).

It is highly desirable for $\mathcal{R}$ to operate near the Nyquist rate, which motivates minimizing OF.
Additionally, $\mathcal{R}$ should be resilient to noise and computationally efficient.
Hence, we focus on developing recovery algorithms that ensure high-speed performance, robustness to noise, and operate at a low sampling rate.

\subsection{Problem Formulation}
\label{Sec:Problem_Formulation}
The problem of recovering $f(n)$ from $f_\lambda(n)$ is equivalent to recovering $z(n)$ from the same $f_\lambda(n)$. Hence, algorithms proposed in this paper focus on recovering $z(n)$.
Specifically, we use the following
two properties, called time-limited-ness and BL-ness, that are associated with $z(n)$.

\textbf{Time-limited-ness \cite{B2R21, Cheb}:} This property states that $z(n)$ is a finite $N$-length signal. 
Let $E$ be the energy of $f(t)$, i.e.,
$\int_{-\infty}^{\infty}|f(t)|^2 dt= E$.
Now consider $T$-duration of $f(t)$ such that
\begin{equation}
\int_{0}^{T}|f(t)|^2 dt = E-(\lambda-\epsilon)^2,
\label{Find_N}
\end{equation}
where $\epsilon>0$ is a small constant value.
This choice of $T$ guarantees that 
$|f(t)|<\lambda,\ \forall t\notin[0, T]$. This implies, $|f(n)|<\lambda,\ \forall n\notin [0, N-1]$, where $N = \lceil\frac{T}{T_s}\rceil$. 
Therefore, 
$f_\lambda(n) = f(n),\ \forall n\notin [0, N-1]$, which implies that $z(n)$ is a finite $N$-length signal and $z(t)$ is a finite $T$-duration signal.

In practice, the signal $f(t)$ can be recorded directly from the analog system for a duration $T$ determined by choosing $\epsilon$ for a fixed $\lambda$, ensuring that $T$ captures 99\% of the total energy $E$.
Alternatively, the $N$-value can be determined in two steps:
\begin{enumerate}
    \item Calculate the energy $E_\lambda$ of $f_\lambda(n)$.
    \item Find $N$ such that 99\% of $E_\lambda$ is included in the signal.
\end{enumerate}

Observe that here we chose the value of $N$ based on \eqref{Find_N}. However, the true support set of $z(n)$, denoted as $\mathcal{N}_\lambda \subset [N-1]$, where  $f(n) = {f_\lambda}(n), \forall\ n{\notin}\mathcal{N}_\lambda$,  can be significantly smaller than $N$. For example, the signal shown in Fig. \ref{Typical_Modulo_Signal} (a) has $N=1024$. But, the true support $\mathcal{N}_\lambda$ is between $(360, 670)$, that is, in practice, $N\gg\mathcal{N}_\lambda$.

\textbf{BL-ness \cite{B2R21}:} This property paves the way to compute the partial frequency spectrum of $z(n)$ and $\hat{z}(n)$.
As $f(t)\in \mathcal{B}_{f_m}$, the sequence $f(n)$ is a BL signal with bandwidth $[-\rho\pi, \rho\pi]$ or $[0, \rho\pi]\cup[2\pi-\rho\pi, 2\pi)$, where $\rho = \frac{2f_m}{f_s} = \frac{1}{\text{OF}}$.
Hence, from \eqref{Modulo_Equation_Discrete}, the frequency spectrum of $z(n)$ is equal to the frequency spectrum of $f_\lambda(n)$ over $[-\pi, -\rho\pi)\cup(\rho\pi, \pi]$ or $(\rho\pi, 2\pi-\rho\pi)$.

Building on the aforementioned two properties, we now proceed to formulate the problem.
Applying the first-order difference operator on \eqref{Modulo_Equation_Discrete} and then utilizing the time-limitedness property, we obtain the following expression:
\begin{equation}
\hat{f}_\lambda(n) = 
    \begin{cases}
      \hat{f}(n) + \hat{z}(n), & \text{if} \ n\in[0, N-1] \\
      \hat{f}(n), & \text{otherwise}
    \end{cases}.
\label{first-order-difference}
\end{equation}
Since $\hat{f}(n)$ is also a BL signal with bandwidth $[-\rho \pi, \rho\pi]$, we have the following Discrete Fourier Transform (DFT) relation for $\hat{z}(n)$: 
\begin{equation}    \hat{F}_\lambda\left(e^{\frac{j2\pi k}{N}}\right) = \sum_{n=0}^{N-1}\hat{z}(n)e^{\frac{-j 2\pi kn}{N}} = \hat{Z}(k),
\end{equation}
where  $k\in U_N = \{p\ |\ p\subset[N-1]\And\frac{2\pi p}{N}\in (\rho\pi, 2\pi-\rho\pi)\}$, $\hat{F}_\lambda\left(e^{j\omega}\right)$ and $\hat{Z}(k)$ denote Discrete-Time Fourier Transform (DTFT) and DFT coefficients of $\hat{f}_\lambda(n)$ and $\hat{z}(n)$, respectively.
Representing the above equation in matrix form leads to
\begin{equation}    \left[\mathbf{\hat{F}}_\lambda\right]_{M\times 1} = [\mathbf{V}]_{M \times N}[\mathbf{\hat{z}]}_{N\times 1},
    \label{eqnToSolve}
\end{equation}
where $M = \#U_N$, $M\ll N$, $\mathbf{V}_{k,n} = e^{\frac{-j2\pi kn}{N}}$, $k\in U_N$, and $n\in[N-1]$. 

Our aim is to estimate $\mathbf{\hat{z}}$ from the observation vector $\mathbf{\hat{F}}_\lambda$ by solving the linear inverse problem in \eqref{eqnToSolve}.
Once the estimation of $\mathbf{\hat{z}}$ is obtained, $\mathbf{z}$ is computed using the cumulative summation operator: ${z}(n) = \sum_{m=0}^n \hat{z}(m)$, thereby facilitating the estimation of $f(n)$ using \eqref{Modulo_Equation_Discrete}.
Two recovery algorithms have been proposed  \cite{Fourier-Prony, B2R2, B2R21} that aim to recover $\mathbf{z}$ from its partial Fourier transform coefficients. 
In \cite{Fourier-Prony}, the authors introduced a spectral estimation method named Fourier-Prony, designed to recover $\mathbf{\hat{z}}$ utilizing the frequency spectrum of $\hat{f}_\lambda(n)$.
However, the Fourier-Prony algorithm requires prior knowledge of the number of folding instants information, which is not readily available unless the modulo system itself generates this additional information.
In the works \cite{B2R2} and \cite{B2R21}, the authors proposed the B$^2$R$^2$ recovery algorithm, which exploits the two aforementioned properties corresponding to \eqref{Modulo_Equation_Discrete}. However, their algorithm relies on the known prior value of $\mathcal{N}_\lambda$.
Moreover, B$^2$R$^2$ formulates a constrained optimization problem to solve \eqref{eqnToSolve}, which estimates $\mathbf{z}$ sample by sample, leading to potential computational inefficiency. 
Hence, there is a need for a recovery algorithm that is fast, robust, and operates at a lower sampling rate. 

The aforementioned formulation does not consider the effect of quantization, which is unavoidable in practice. With quantization, the folded samples are denoted by $f_\lambda^q(n)$. Typically, under high-resolution quantization assumptions, the effect of quantization can be modeled as an additive noise term. Mathematically, we have
\begin{align}
    f_\lambda^q(n) = f_\lambda(n) + \epsilon_q(n), \label{eq:quant_noise}
\end{align}
where $\epsilon_q(n)$ is quantization noise. Noting that $|f_\lambda(n)| \leq \lambda$, a $B$-bit quantization ensures that $|\epsilon_q(n)|\leq \frac{\Delta_B}{2}$ where $\Delta_B = \frac{2\lambda}{2^B}$ is the quantization step size. In the additive quantization model, $\epsilon_q(n)$ can be considered as i.i.d. random variable with a uniform distribution between the interval $[-\Delta_B/2, \Delta_B/2]$. 

A larger $B$ results in a lower quantization error. However, this increases the memory requirement, computational cost, and the bit rate. The latter is particularly crucial in modulo-folding-based ADC, which operates above the Nyquist rate. For a given bit budget of $B$ and sampling rate, is it possible to reduce the reconstruction error incurred in the framework mentioned earlier? We show that this is possible by \textit{bits distribution}. In particular, we use one bit out of the available $B$ bits to quantify the folding instants. 
From \eqref{Modulo_Equation_Discrete}, $z(n)$ is a piece-wise constant residual signal whose values are integer multiple of $2\lambda$. A $2\lambda$ level jump in $z(n)$ denotes the occurrence of folding. Let $b(n)$ denote the bit-stream that encodes information about the folding instants. Whenever $z(n)$ changes its amplitude, $b(n)$ is set as 1; otherwise, zero.  In this way, the one-bit encodes the folding instants, and the remaining bits are used to quantize the folded samples. We are not concerned with the exact time instant or value of the $2\lambda$ level jump in $z(n)$.
We show that the unfolding algorithm will be a simple least-squares with the \textit{bits distribution} mechanism. We demonstrate that this approach can be implemented in hardware without requiring any additional circuitry beyond an OR gate in the existing modulo-ADC architecture.

In the following section, we first discuss the proposed LASSO-B$^2$R$^2$ algorithm and then discuss the \textit{bits distribution} mechanism and its hardware implementation details in the subsequent section.

\section{LASSO-B$^2$R$^2$}
\label{LASSO_B2R2}
This section begins with a crucial observation, which states that $\hat{\mathbf{z}}$ is a sparse vector. 
To support this claim, we derive an upper bound on the number of non-zero elements in $\hat{\mathbf{z}}$. 
Based on this observation, we then formulate the problem of recovering $\mathbf{\hat{z}}$ from $\mathbf{\hat{F}}_\lambda$ as a sparse recovery problem, which we solve using the well known Iterative Soft Thresholding Algorithm (ISTA) \cite{Eldar, LIP2}.

\subsection{On the Sparsity of $\mathbf{\hat{z}}$}
\label{Observation}
Consider the vector $\mathbf{\hat{z}}$, where a non-zero value at a specific index signifies a level-crossing at $(2\mathbb{Z}+1)\lambda$ in the signal $f(t)$ or a level jump of $2\lambda\mathbb{Z}$ in $z(t)$.
The following theorem provides an upper bound on the number of level-crossings or level jumps:
\begin{theorem}
Let $f(t) \in L^2 (\mathbb{R})\cap \mathcal{B}_{\omega_m},\ t\in\mathbb{R}$.
The corresponding discrete modulo samples are $f_{\lambda}(n) = f(n)+z(n), n\in\mathbb{Z}$, where $f_\lambda(n) = f(n),\ \forall n\notin [0, N-1]$, implying that $z(n)$ is a finite $N$-length signal. Let $L$ denote the number of non-zero elements in $\hat{z}(n)$. Then, $L$ satisfies the following inequality:
\begin{equation}
L \leq \text{min}\left(4K + 4K\Bigg\lfloor\frac{||f(t)||_\infty-\lambda}{2\lambda}\Bigg\rfloor, N\right),
\label{upperbound}
\end{equation}
where $K  = \lfloor\frac{N}{2\text{OF}}\rfloor$.
\end{theorem}
\begin{proof}
To establish an upper bound for $L$, we begin by truncating $f(t)$ to a duration $T$, given by $T = NT_s$, and denoted as $\Tilde{f}(t)$. As $f(t)$ has bandwidth $\mathcal{B}_{\omega_m}$, the truncated signal $\Tilde{f}(t)$ lies within $\mathcal{B}_{(\omega_m+\Delta \omega)}$. Here, $\Delta \omega$ accounts for the spectral leakage introduced during the truncation process. Beyond the duration $T$, $f(t)$ possesses negligible energy and can be represented as ${f}_\lambda(t)$. Consequently, we approximate the bandwidth of $\Tilde{f}(t)$ to be $[-\omega_m, \omega_m]$.
We then formulate a $T$-periodic signal, defined as:
\begin{equation*}
    f_p(t) = \sum_{l{\in}\mathbb{Z}}^{}\tilde{f}(t-lT).
\end{equation*}
The Fourier series representation of $f_p(t)$, denoted by $F_p(k)$, is expressed as:
\begin{equation}
f_p(t) = \sum_{k = -K}^{K} F_p(k) e^{-jk\omega_0t},
\label{TrigPoly}
\end{equation}
where $\omega_0 = \frac{2\pi}{T}$ and $K = \lfloor\frac{\omega_m}{\omega_0}\rfloor = \lfloor\frac{N}{2\text{OF}}\rfloor$.
The above equation represents a trigonometric polynomial of order $K$. For a trigonometric polynomial $q(t)$ of order $K$ over a period $T$, it is known that it can have a maximum of $2K$ level-crossings for a level $l$, where $|l| \leq ||q(t)||_\infty$ \cite{powellapproximation}.
Further, the number of $(2\mathbb{Z}+1)\lambda$ level-crossings, referred to as folding levels, in $f(t)$ is given by:
\begin{equation*}
    2 + 2\Big\lfloor\frac{||f_p(t)||_\infty-\lambda}{2\lambda}\Big\rfloor.
\end{equation*}
Consequently, an upper bound $\Tilde{L}$ for the number of times $f_p(t)$ crosses the $(2\mathbb{Z}+1)\lambda$ levels can be expressed as:
\begin{equation}
\Tilde{L} \leq \left(2 + 2\Big\lfloor\frac{||f_p(t)||_\infty-\lambda}{2\lambda}\Big\rfloor\right)2K.
\end{equation}
This leads to the conclusion that the number of $(2\mathbb{Z}+1)\lambda$ level-crossings in $f(t)$ or the level jumps of $2\lambda\mathbb{Z}$ in $z(t)$ is bounded by:
\begin{equation}
\Tilde{L} \leq 4K + 4K\Bigg\lfloor\frac{||f(t)||_\infty-\lambda}{2\lambda}\Bigg\rfloor.
\end{equation}
Considering the above insights and the fact that $\mathbf{\hat{z}}$ is an $N$-length vector, the inequality in \eqref{upperbound} follows.
\end{proof}

\begin{table}[t!]
\centering
\caption{\small{Upper bound on L for different $\lambda$ and OF values.}}
\footnotesize
\begin{adjustbox}{max width=\textwidth}
\renewcommand{\arraystretch}{1.5}
\scalebox{0.8}{
\begin{tabular}{|M{0.7cm}|M{0.8cm}|M{1.2cm}|M{0.8cm}|M{1.2cm}|M{0.8cm}|M{1.2cm}|} \hline
\bfseries{\small{}} & 
\multicolumn{2}{|c|}{\bfseries{\footnotesize{OF = 4}}} & 
\multicolumn{2}{|c|}{\bfseries{\footnotesize{OF = 6}}} & 
\multicolumn{2}{|c|}{\bfseries{\footnotesize{OF = 8}}} \\ \cline{2-7}
\bfseries{$\lambda$} & \bfseries{\footnotesize{$2K$}} & \bfseries{\footnotesize{Upper bound on $L$}} & \bfseries{\footnotesize{$2K$}} & \bfseries{\footnotesize{Upper bound on $L$}} & \bfseries{\footnotesize{$2K$}} & \bfseries{\footnotesize{Upper bound on $L$}} \\ \hline 
\small{$0.75$} &	$256$	&	$512$ &	$170$	&	$340$ &	$128$	&	$256$	\\	\hline
\small{$0.5$} &	$256$	&	$512$ &	$170$	&	$340$ &	$128$	&	$256$	\\	\hline
\small{$0.25$} & $256$	&	$1024$ & $170$	&	$680$ & $128 $ & $512$ \\    \hline
\small{$0.05$} &	$256$	&	$1024$ & $170$	&	$1024$ & $128 $ & $1024$	\\	\hline
\end{tabular}}
\end{adjustbox}
\label{tab:Upper bound on L for  different cases}
\normalsize
\end{table}
\begin{figure}[t!]
    \centering
    \includegraphics[height = 4.7cm, width = 9.5cm]{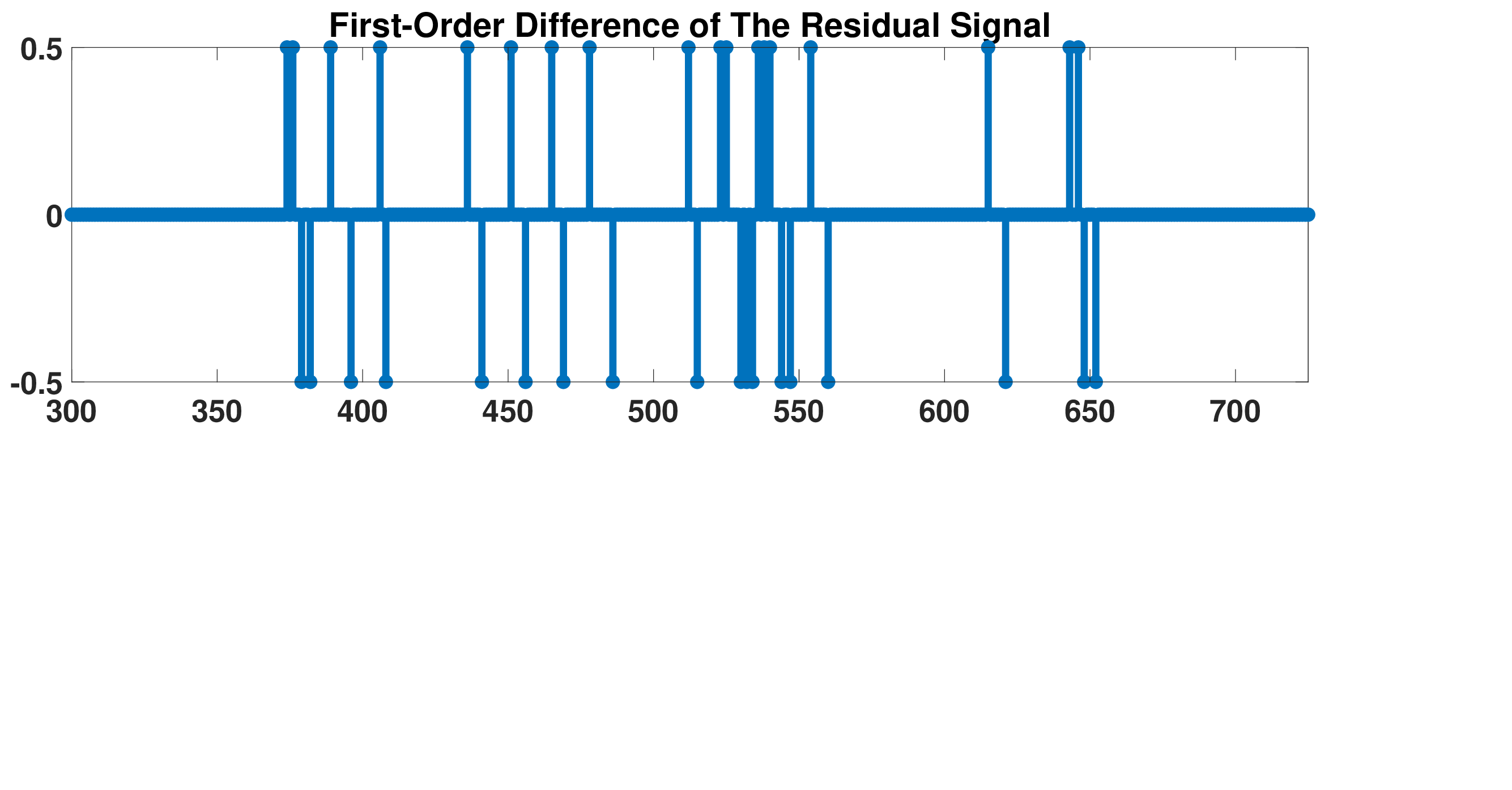}
    \vspace{-2.2cm}
    \caption{\small{First-order difference of the residual signal, shown in Fig. \ref{Typical_Modulo_Signal}(b).}}
    \label{Fig_First_Order_Diff}
\end{figure}
As an example, assume that the maximum value of $f(t)$ is normalized to $1$ and $N = 1024$. Table \ref{tab:Upper bound on L for different cases} shows the upper bound on $L$ for different $\lambda$ values of a particular OF.
From Table \ref{tab:Upper bound on L for different cases}, the following insights are drawn: For moderate $\lambda$ and OF values $\mathbf{\hat{z}}$ exhibits sparsity. 
A reduced $\lambda$ value increases the bound on $L$, while a higher OF value diminishes the bound.
In practice, the actual value of $L$ is often significantly lower than the theoretical upper bound. Fig. \ref{Fig_First_Order_Diff} shows the first-order difference of the residual signal depicted in Fig. \ref{Typical_Modulo_Signal}(b) with $N=1024$, $\text{OF}=6$, and $\lambda = 0.25$. In this case, the actual value of $L$ is $36$, which is significantly lower than the theoretical upper bound of $680$.
This substantial difference can be intuitively explained by the fact that our theoretical analysis is based on $N$, while the actual support of $\mathbf{z}$ is $\mathcal{N}_\lambda$, which is considerably smaller than $N$.

\subsection{Algorithm}
Based on the observed sparsity of $\mathbf{\hat{z}}$, we address the problem of estimating $\mathbf{\hat{z}}$ in  \eqref{eqnToSolve} by formulating it as a 
sparse recovery problem
\begin{equation}
    {\argmin_\mathbf{\hat{z}}}\  \frac{1}{2}||\mathbf{\hat{F}}_\lambda - \mathbf{V}\mathbf{\hat{z}}||_2^2 + \gamma||\mathbf{\hat{z}}||_1,
    \label{LASSO}
\end{equation}
where $\gamma$ serves as the regularization parameter. Numerous algorithms have been developed to solve problem \eqref{LASSO}. 
Here, we employ the widely recognized ISTA \cite{ProximalAlg} method. The update of $\mathbf{\hat{z}}$ at the $(i+1)^{th}$ iteration is given by:
\begin{equation*}
    \mathbf{\hat{z}}^{(i+1)} = S_{\gamma\tau}\left(\mathbf{\hat{z}}^{(i)}-\tau\mathbf{V}^H\left(\mathbf{V}\mathbf{\hat{z}}^{(i)} - \mathbf{\hat{F}}_\lambda\right)\right),
\end{equation*}
where $\tau$ is a step-size and $S_{\gamma\tau}(.)$ is the soft-thresholding operator,
\begin{equation*}
    S_{\gamma\tau}(x) = \text{sign}(x)\text{max}\left(|x|-\gamma\tau,0\right),\ x\in\mathbb{R}.
\end{equation*}
As suggested in \cite{ProximalAlg}, for convergence we initialize $\gamma = 0.1 ||\mathbf{V}^H\mathbf{\hat{F}}_\lambda||_\infty$ and $\tau = \frac{1}{||\mathbf{V}||_2^2}$.
Similar to B$^2$R$^2$, this work leverages beyond-the-bandwidth frequency information to reconstruct the residual samples through a LASSO problem formulation. Consequently, we refer to this algorithm as LASSO-B$^2$R$^2$. A detailed summary of the proposed algorithm is outlined in Algorithm $1$.
\begin{algorithm}
\caption{LASSO-B$^2$R$^2$}\label{alg:LASSO-B2R2}
\begin{algorithmic}[1]
\State \textbf{Input:} $f_\lambda(n)$, $\lambda$, $\rho$, maxItr \Comment{maxItr denotes the maximum number of iterations}
\State Compute $N$ using \eqref{Find_N}
\State Find $k$ values, where $k\in U_N = \{p\ |\ p\subset[N-1]\And\frac{2\pi p}{N}\in (\rho\pi, 2\pi-\rho\pi)\}$
\State Compute $M$, where $M=\# U_N$
\State Construct $[\mathbf{V}]_{M\times N}$,
where $V(k,n) = e^{\frac{-j2\pi kn}{N}}$, $n \in[N-1]$, and $k\in U_N$
\State Compute ${\hat{F}}_\lambda(e^{\frac{j2\pi k}{N}})$, $\forall$ $k\in U_N$ 
\State \textbf{Initialize:} $\gamma = 0.1 ||\mathbf{V}^H\mathbf{\hat{F}}_\lambda||_\infty$, $\tau = \frac{1}{||\mathbf{V}||_2^2}$, maxItr = $1000$, and $\mathbf{\hat{z}}^{(0)}\in \mathcal{N}(0,1)$
\For{$i=0$ : maxItr}
    \State $\mathbf{\hat{z}}^{(i+1)} = S_{\gamma\tau}\left(\mathbf{\hat{z}}^{(i)}-\tau\mathbf{V}^H\left(\mathbf{V}\mathbf{\hat{z}}^{(i)} - \mathbf{\hat{F}}_\lambda\right)\right)$
    \If{$||\mathbf{\hat{z}}^{(i+1)} - \mathbf{\hat{z}}^{(i)}||_2 < 10^{-4}$} 
        \State $\mathbf{\hat{z}} = \mathbf{\hat{z}}^{(i+1)}$
        \State exit
    \EndIf
\EndFor
\State $\mathbf{\hat{z}} \gets \lceil\frac{\lfloor \mathbf{\hat{z}}/\lambda\rfloor}{2}\rceil$  \Comment{Rounding the residual to $2\lambda\mathbb{Z}$}
\State $\mathbf{z}\gets \text{cumsum}(\mathbf{\hat{z}})$ \Comment{Cumulative summation operator on $\mathbf{\hat{z}}$}
\State $f(n)\gets f_\lambda(n)-z(n)$
\State \textbf{Output:} $f(n)$
\end{algorithmic}
\end{algorithm}
\begin{figure*}[t!]
    \centering
    \includegraphics[height = 8cm, width = 15cm]{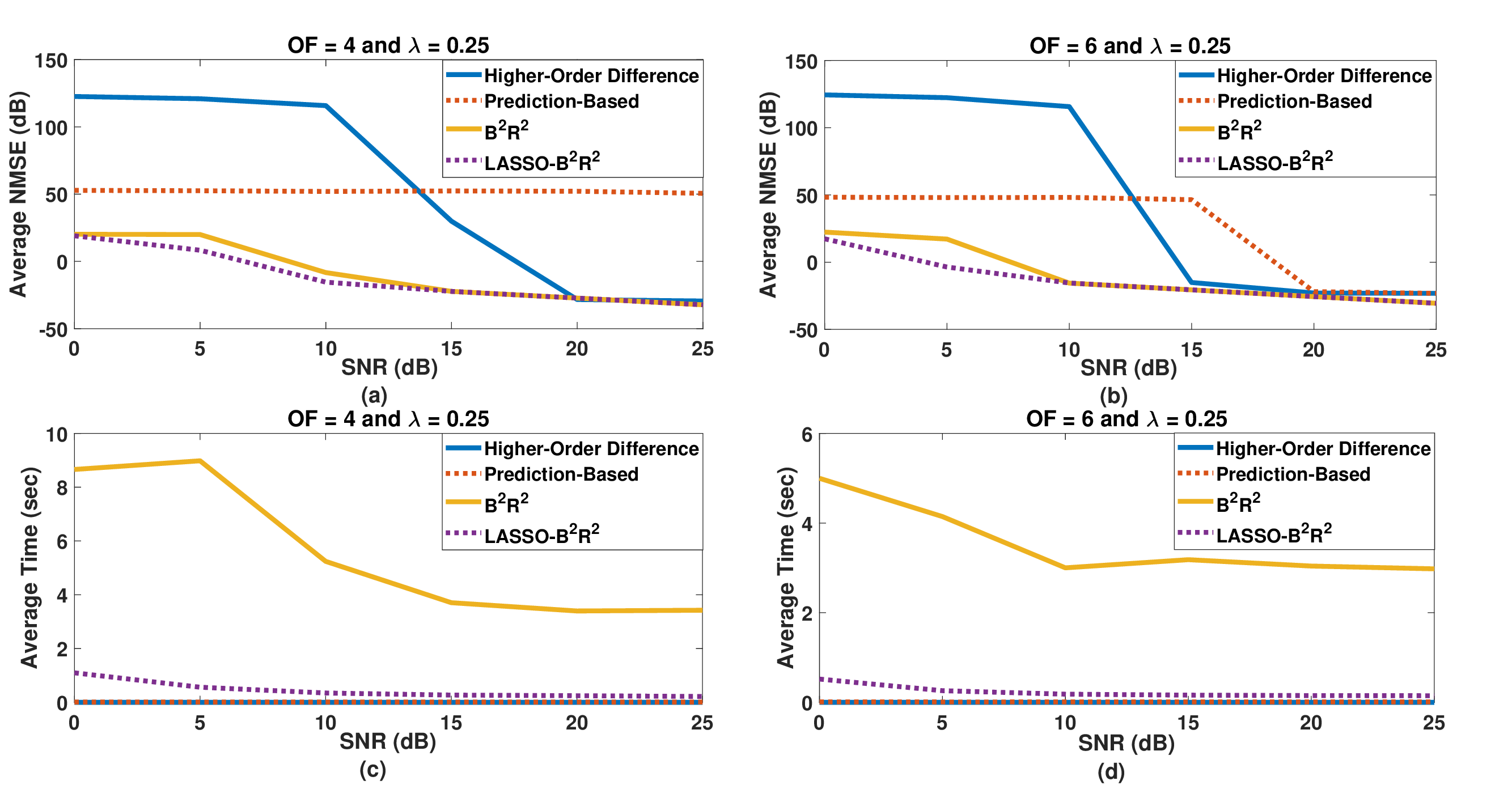}
    \caption{\small{Average NMSE versus SNR  and average time versus SNR for OF$=4$ and OF$=6$ with $\lambda=0.25$.}}
    \label{fig:2}
\end{figure*}
\subsection{Simulations}
In this simulation, we evaluate the performance of LASSO-B$^2$R$^2$ against HOD-based \cite{US2}, prediction-based \cite{Cheb}, and B$^2$R$^2$ \cite{B2R2} algorithms.  The simulations were conducted using Python version 3.9 on an AMD Ryzen 7 3700x 8-core processor with 16 GB RAM.
\begin{figure}[t!]
    \centering
    \includegraphics[height = 8cm, width = 16.5cm]{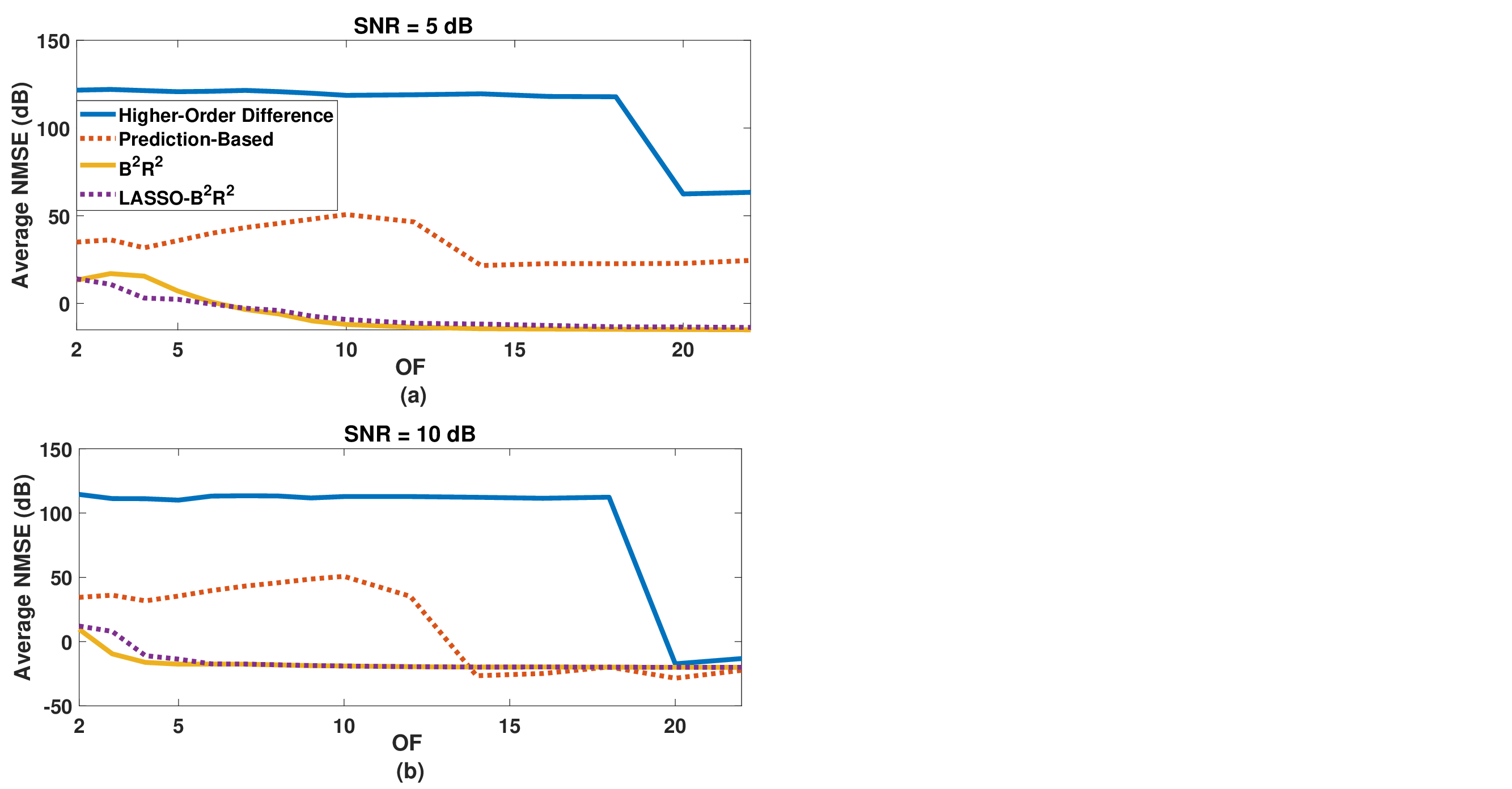}
    \caption{\small{Average NMSE versus OF for SNR$=5$ dB and SNR$=10$ dB.}}
    \label{fig:3}
    \vspace{-0.2cm}
\end{figure}

Initially, we assess their robustness. Therefore, we add noise with a particular Signal-to-Noise Ratio (SNR) value to $f_\lambda(n)$, characterized by specific OF and $\lambda$ values. The noise adheres to a Gaussian distribution with a mean of zero and a variance denoted as $v$, calculated based on the provided SNR value. Subsequently, we employ the various algorithms to estimate $f(n)$.
Let $\bar{f}(n)$ represent the estimated unfolded signal obtained through a specific algorithm. The Normalized Mean Square Error (NMSE) is given by:
\begin{equation*}
\text{NMSE } = \frac{||f(n)-\bar{f}(n)||_2^2}{||f(n)||_2^2}. 
\end{equation*}
To compute the average NMSE, we repeat this calculation across $250$ i.i.d noise instances.
Fig. \ref{fig:2}(a) and Fig. \ref{fig:2}(b) illustrate the average NMSE as SNR varies from $0$ to $25$ dB for OF$ =4$ and OF$ = 6$, respectively. Similarly, Fig. \ref{fig:2}(c) and Fig. \ref{fig:2}(d) show the average (over $250$ noise instances) time taken by each algorithm against SNR to estimate the true unfolded signal.
Here, we consider signals with $N=1024$, and each signal's maximum amplitude is normalized to $1$.
From Fig. \ref{fig:2}, we conclude that the LASSO-B$^2$R$^2$ algorithm exhibits greater robustness compared to the HOD-based and prediction-based methods, albeit at a slightly higher computational cost. The robustness of LASSO-B$^2$R$^2$ is on par with B$^2$R$^2$, but with reduced computational time.
As mentioned earlier, in the B$^2$R$^2$ approach, we recover the ${z(n)}$ samples individually. Specifically, estimating a sample involves solving a constrained optimization problem using an iterative algorithm; this process is repeated for all samples in $z(n)$. In contrast, the LASSO-B$^2$R$^2$ method recovers the entire residual signal at once.

Next, we evaluate the performance of LASSO-B$^2$R$^2$ in relation to the required OF value. We focus on the modulo signal at a specific SNR value with $\lambda = 0.25$. For each OF value tested, we generated $250$ signal realizations and calculated the average NMSE.
Fig. \ref{fig:3}(a) and Fig. $\ref{fig:3}$(b) display the average NMSE results across OF values ranging from $2$ to $22$, corresponding to SNR values of $5$ dB and $10$ dB, respectively. Based on these findings, we infer that LASSO-B$^2$R$^2$ demands a lower OF value compared to methods relying on higher-order differences and polynomials. However, its performance is comparable to that of B$^2$R$^2$.

\subsection{Discussion on Number of Measurements:}
\begin{figure}[h]
    \centering
    \begin{subfigure}[b]{0.45\textwidth}
        \centering
        \includegraphics[height = 6cm, width = 14cm]{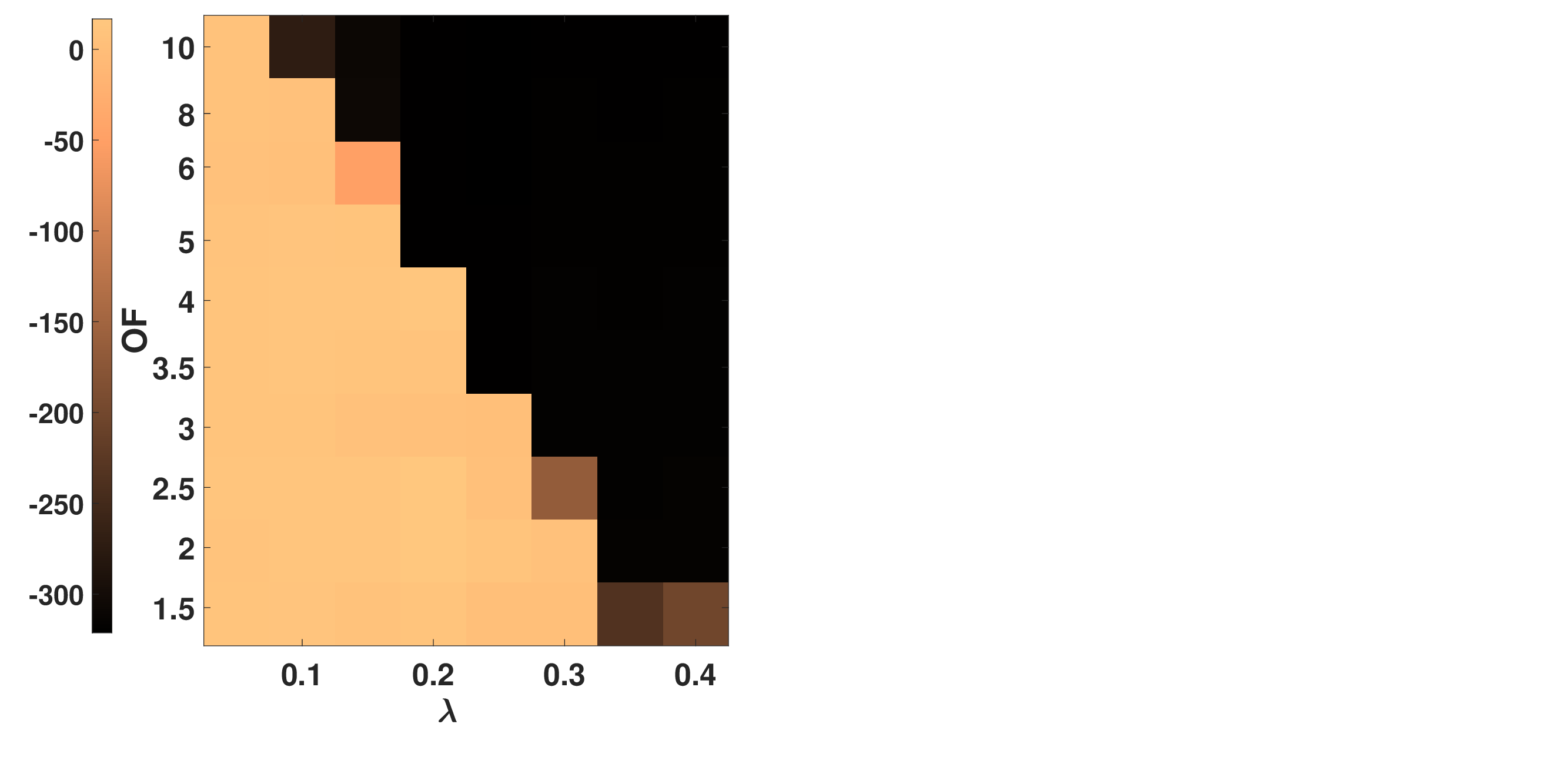}  
        \caption{}
        \label{fig:4a}
    \end{subfigure}
    \hfill
    \begin{subfigure}[b]{0.45\textwidth}
        \centering
    \includegraphics[height = 6cm, width = 14cm]{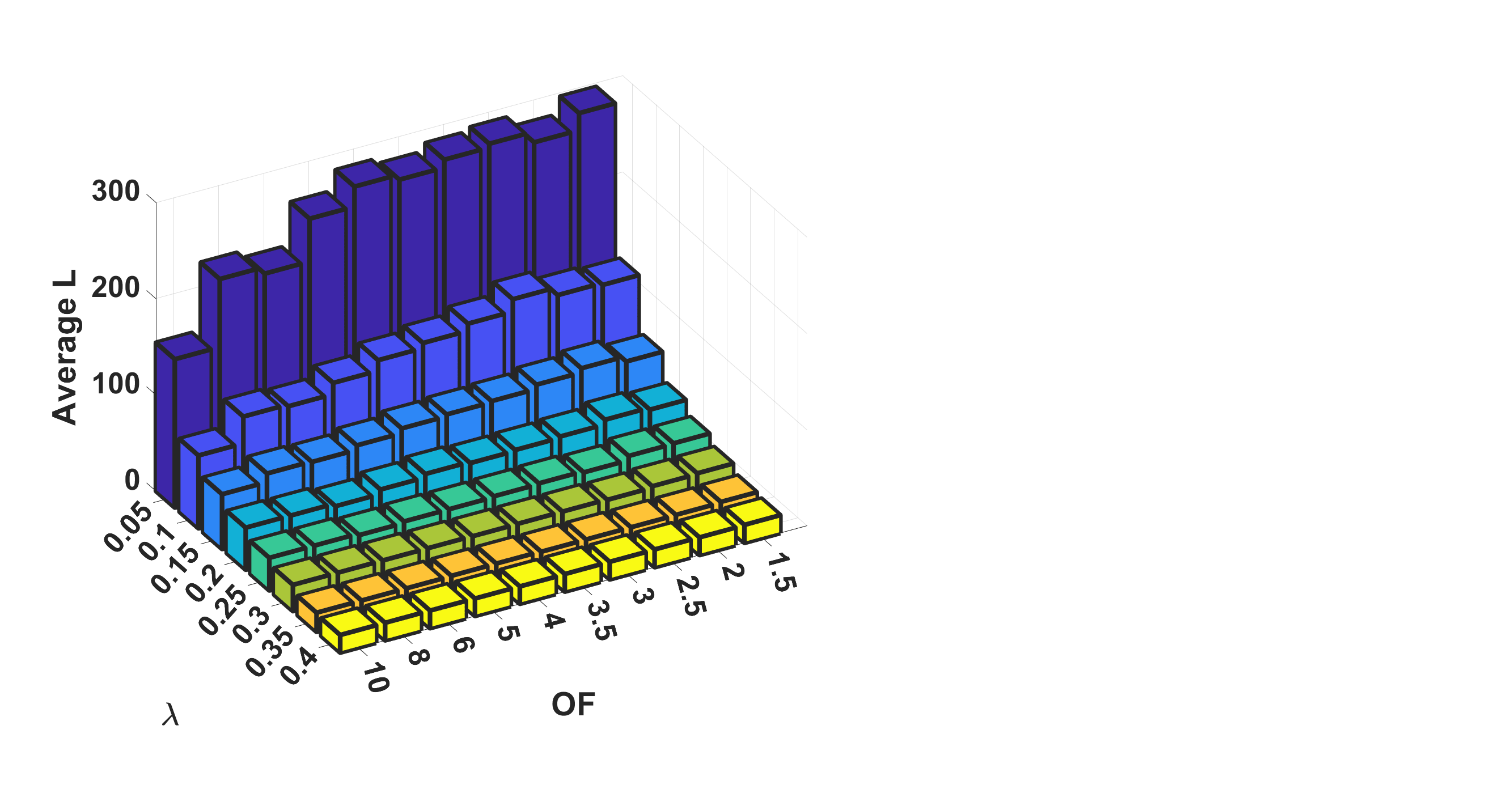}  
        \caption{}
        \label{fig:4b}
    \end{subfigure}
    \caption{\small{(a)-(b) Average NMSE and average $L$ for different OF and $\lambda$ pair values, respectively.}}
    \label{fig:4}
\end{figure}

Here, we discuss the relation between the number of frequency measurements $M$ in \eqref{eqnToSolve} and the sparsity of $\mathbf{\hat{z}}$, which are controlled by the parameters OF and $\lambda$, respectively.
Since $\mathbf{V}$ is a partial DFT matrix, the \textit{spark} of $\mathbf{V}$ is at least $M+1$. Thus, the spark-based guarantee provides the following relation between $M$ and $L$:
\begin{equation}
    \|\mathbf{\hat{z}}\|_0 = L < \frac{M+1}{2}.
\end{equation}
This provides an upper bound on the number of $2\lambda\mathbb{Z}$ level jumps in $z(t)$.
In practice, for successful recovery, $L$ should be considerably less than $\frac{M+1}{2}$.
For example, in the analysis above, with $N = 1024$ and $\lambda = 0.25$, the number of discrete frequencies for OF $= 6$ that lie within $[0, \rho\pi] \cup [2\pi - \rho\pi, 2\pi)$ is $170$, i.e., $2K = 170$. This implies $M = N - 2K = 854$. Moreover, for the given signal, $L = 36$, which is significantly less than $\frac{M+1}{2} = 427$. 
As previously mentioned, a smaller $\lambda$ value tends to increase $L$. Therefore, in such scenarios,  a slightly higher sampling rate is required to satisfy the \textit{spark}-based condition; which can be observed from Fig.~\ref{fig:4}. Fig.~\ref{fig:4}(a) and Fig.~\ref{fig:4}(b) illustrate the average NMSE and the corresponding average $L$, respectively, for different combinations of OF and $\lambda$. For each pair of OF and $\lambda$, the average NMSE and average $L$ were computed using 100 signal realizations.

When \(\lambda\) is small, \(L\) becomes large. In such cases, conventional methods like ISTA often fail to recover \(\mathbf{\hat{z}}\) accurately. This challenge can be addressed using Tail-\(\ell_1\)-minimization algorithms.
    
Consider the representation in \eqref{eqnToSolve}, $\mathbf{\hat{F}}_\lambda = \mathbf{V}\mathbf{\hat{z}}$, where \(\mathbf{\hat{z}}\) is assumed to be non-sparse. In this context, aiming to approximate \(\mathbf{\hat{z}}\) by an \(L\)-sparse vector, \(\mathbf{\hat{z}}_L\), using \(\ell_1\)-minimization, \eqref{LASSO}, recovers a signal \(\mathbf{\hat{z}}_1\) such that
\begin{equation}
\| \mathbf{\hat{z}}_1 - \mathbf{\hat{z}} \|_2 \leq C_0 \frac{\| \mathbf{\hat{z}} - \mathbf{\hat{z}}_L \|_1}{\sqrt{L}},
\end{equation}
provided that the \(2L\)-restricted isometry constant of \(\mathbf{V}\) satisfies \(\delta_{2L} < \sqrt{2} - 1\), a condition met by many structured matrices such as partial DFT matrix $\mathbf{V}$ \cite{tail_mot_2, Candes2, Candes1}.
Notably, the $\ell_2$-norm of the recovery error depends on the $\ell_1$-norm of the smallest $N-L$ non-zero entries of \(\mathbf{y}\), the tail part. As a result, recent works have investigated minimizing the tail of the target vector to enhance recovery performance \cite{tail_l1_minimize, tailFISTA, Pradhan}. In particular, it is theoretically shown in \cite{tail_l1_minimize} that, for a full spark matrix, Tail-\(\ell_1\)-minimization can recover \(\mathbf{\hat{z}}\) even when \(L > \frac{M+1}{2}\).
In the Tail-\(\ell_1\)-minimization algorithm, at each \(l\)-th iteration, a support index set \(T\) is chosen based on the indices of the \(L\) largest components of the previous estimate, \(\mathbf{\hat{z}}^{(l-1)}\). The update \(\mathbf{\hat{z}}^{(l)}\) is then obtained by solving:
\[
\mathbf{\hat{z}}^{(l)} = \arg \min_{\mathbf{\hat{z}}} \| \mathbf{\hat{z}}_{T^c} \|_1 \quad \text{s.t.} \quad \mathbf{\hat{F}}_\lambda = \mathbf{V}\mathbf{\hat{z}},
\]
where \(T^c\) includes indices outside \(T\), partitioning \(\mathbf{\hat{z}}\) into subsets \(T\) and \(T^c\). Various algorithms are based on this Tail-\(\ell_1\)-minimization approach \cite{tail_l1_minimize, tailFISTA, Pradhan}. With this motivation, one may formulate the problem of estimating \(\mathbf{\hat{z}}\) as a Tail-\(\ell_1\)-minimization problem for small \(\lambda\) values to improve modulo recovery accuracy.  
However, in the following section we discuss an alternative computationally efficient algorithm to improve the modulo recovery for lower $\lambda$ values.


\section{LASSO-B$^2$R$^2$ With Bits-Distribution}
In this part of the paper, we consider the effect of quantization of the folded samples as discussed in Section~\ref{Sec:Problem_Formulation}. Specifically, we start from the quantized samples as in \eqref{eq:quant_noise} and discuss the proposed \textit{bits distribution}. 

\begin{figure*}[htbp] 
    \centering
    \begin{subfigure}{0.32\textwidth}
        \includegraphics[width=\linewidth]{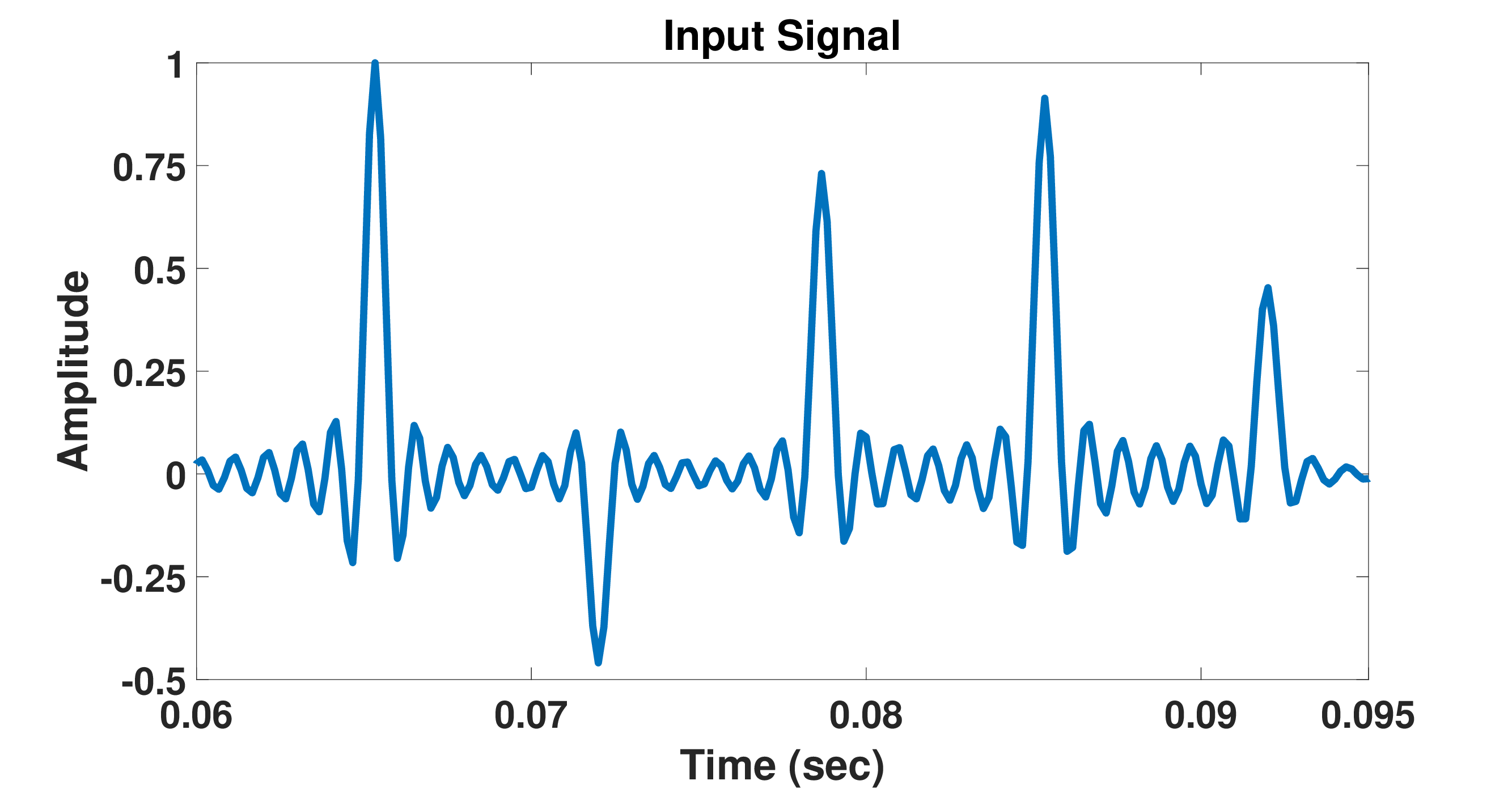} 
        \caption{}
        \label{fig:sub1}
    \end{subfigure}
    \hfill
    \begin{subfigure}{0.32\textwidth}
        \includegraphics[width=\linewidth]{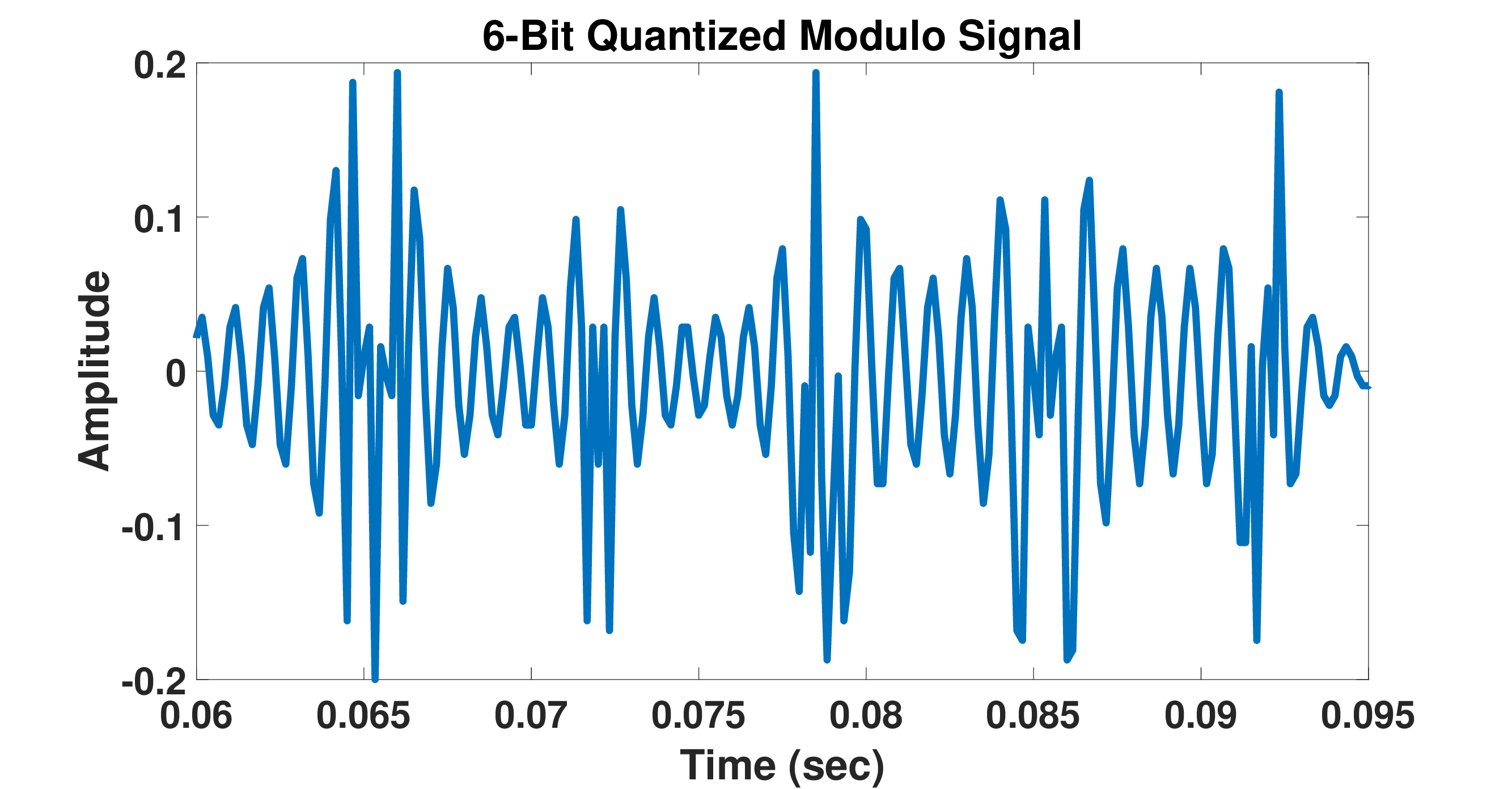}
        \caption{}
        \label{fig:sub2}
    \end{subfigure}
    \hfill
    \begin{subfigure}{0.32\textwidth}
        \includegraphics[width=\linewidth]{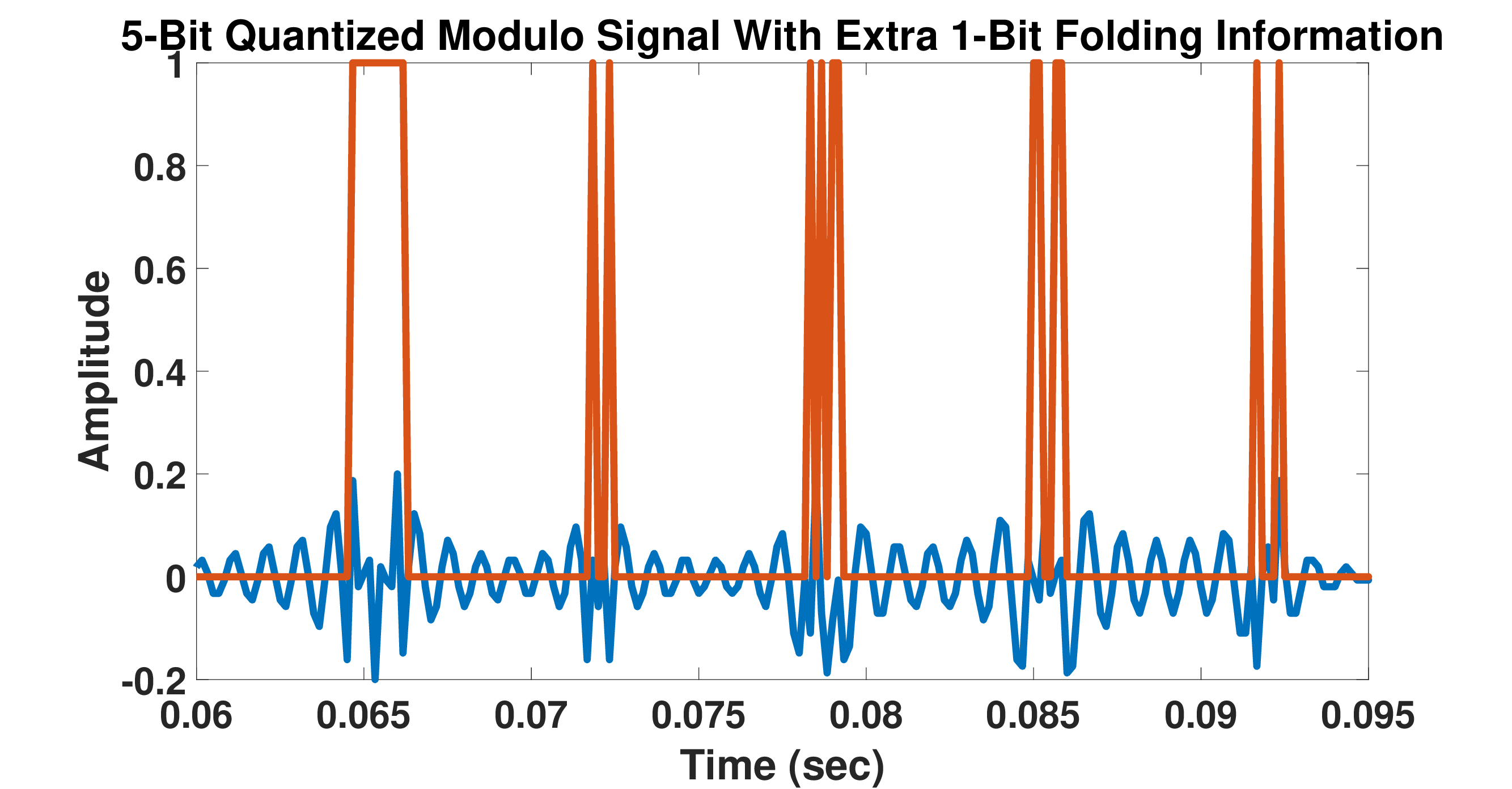}
        \caption{}
        \label{fig:sub3}
    \end{subfigure}
    
    \begin{subfigure}{0.32\textwidth}
        \includegraphics[width=\linewidth]{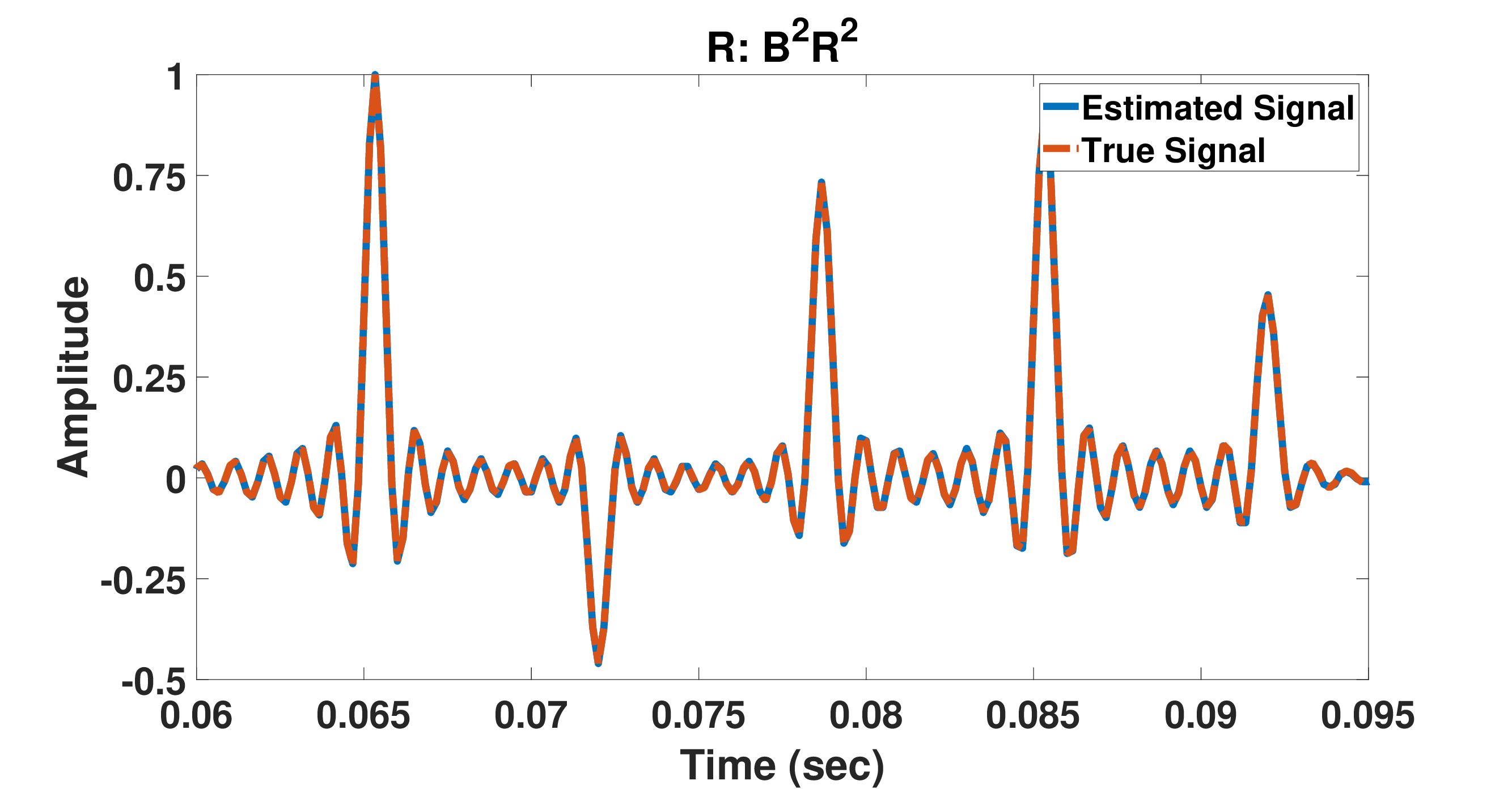}
        \caption{}
        \label{fig:sub4}
    \end{subfigure}
    \hfill
    \begin{subfigure}{0.32\textwidth}
        \includegraphics[width=\linewidth]{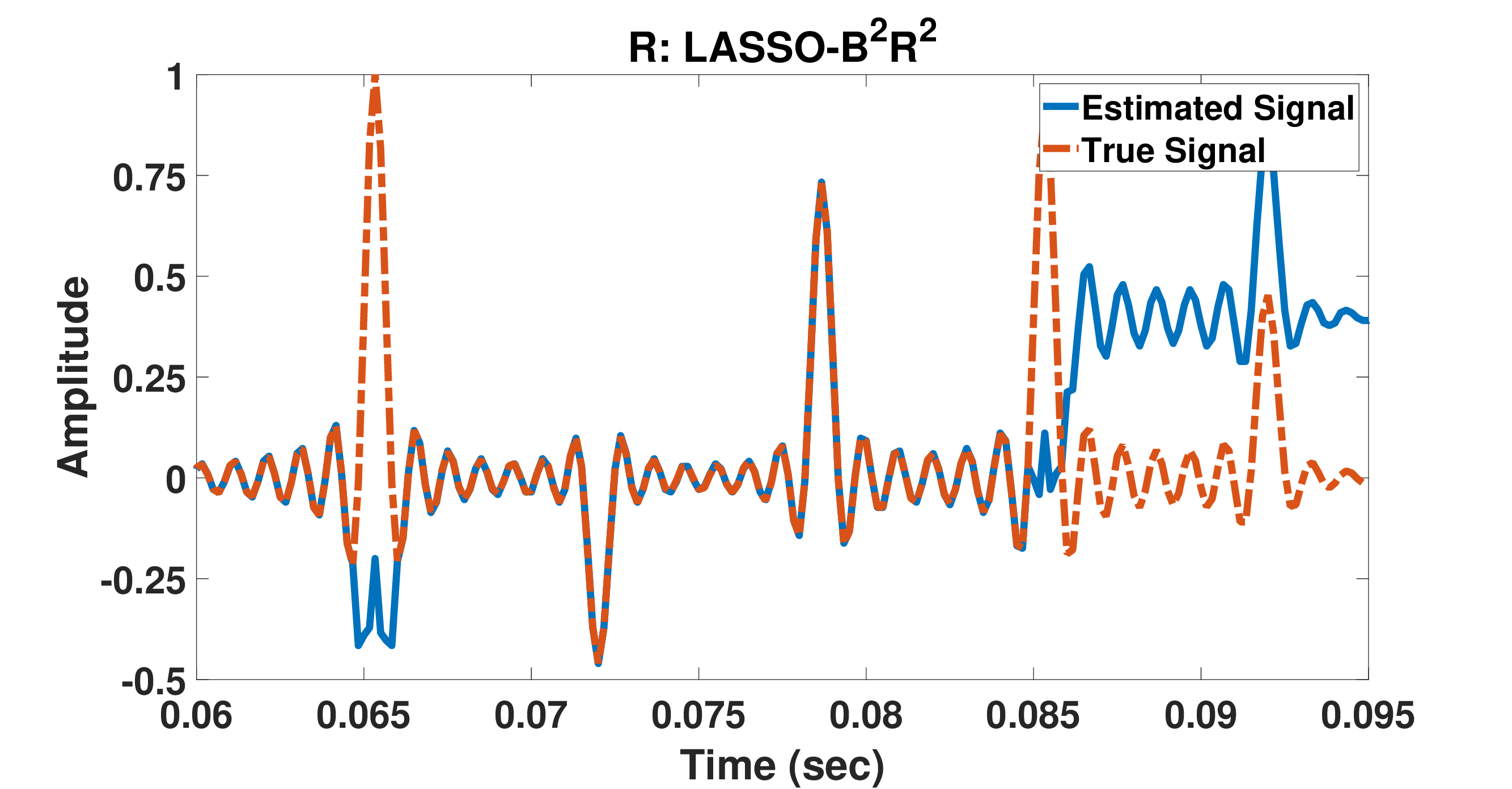}
        \caption{}
        \label{fig:sub5}
    \end{subfigure}
    \hfill
    \begin{subfigure}{0.32\textwidth}
        \includegraphics[width=\linewidth]{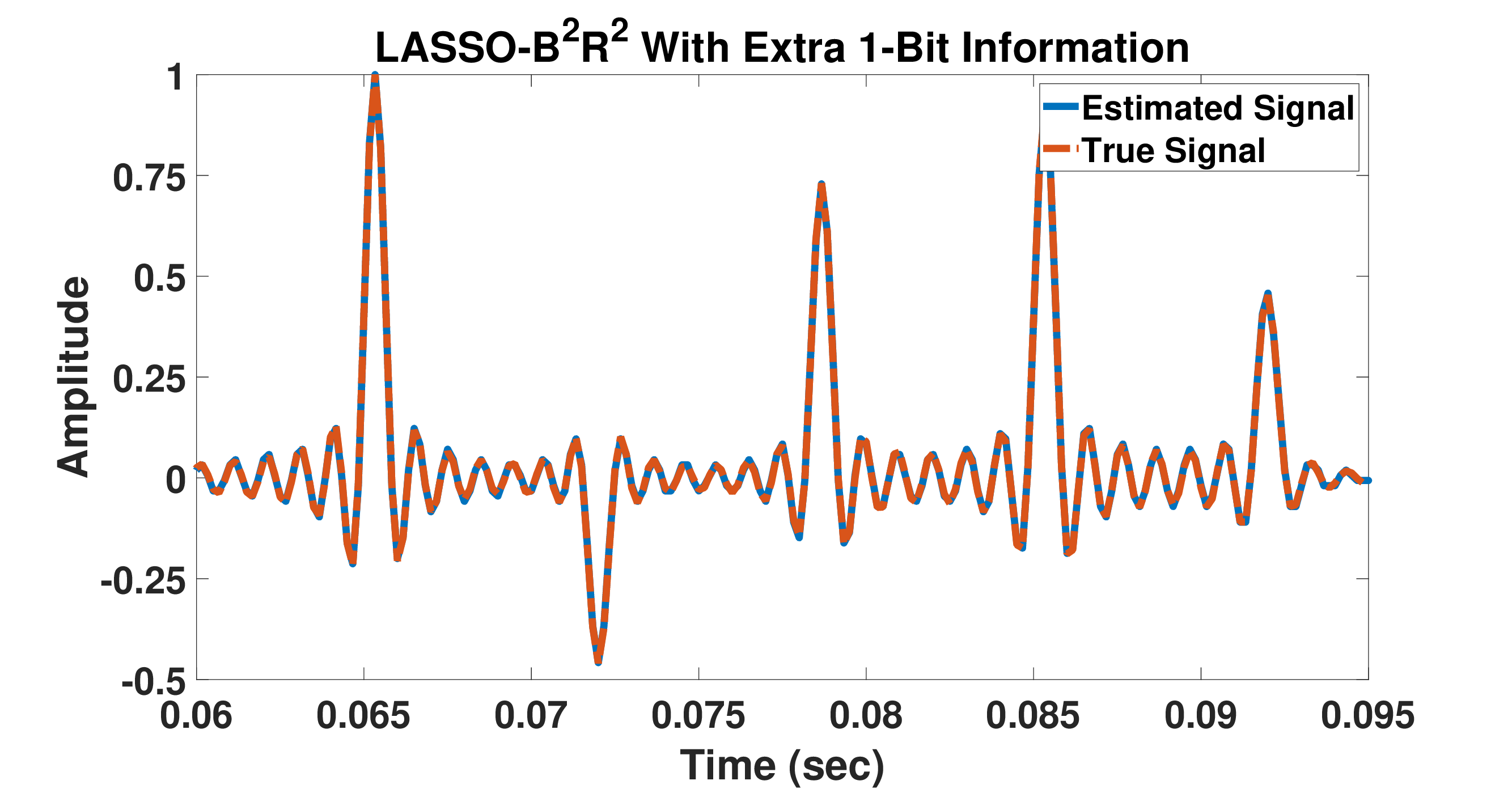} 
        \caption{}
        \label{fig:sub6}
    \end{subfigure}
    
    \caption{\small{(a) BL signal with High DR. (b) Sampled and quantized modulo signal with OF$=3$ and b$=6$. (c) Sampled and quantized modulo signal with OF$=3$ and b$=5$ (blue color); along with the 1-bit information (orange color). (d)-(f) Estimated true signal using B$^2$R$^2$, LASSO-B$^2$R$^2$, and LASSO-B$^2$R$^2$ with 1-bit, respectively.}}
    \label{fig:Comparison}
\end{figure*}
\subsection{Bits-Distribution and The Least-Squares Problem}
Consider the unfolding problem considered in the previous section that amounts to estimating the sparse vector $\mathbf{\hat{z}}$ from the partial Fourier measurements $\mathbf{\hat{F}}_\lambda = \mathbf{V}\mathbf{\hat{z}}$. In the presence of quantization, the Fourier measurements are given as 
\begin{align}
    \mathbf{\hat{F}}_\lambda^q = \mathbf{V}\mathbf{\hat{z}} + \mathbf{w},
    \label{eq:fourier_mes_quant}
\end{align}
where $\mathbf{w}$ is the additive noise term resulting from the quantization of the samples. Note that unlike $\epsilon_q(n)$ in \eqref{eq:quant_noise}, the elements of $\mathbf{w}$ are neither independent nor identically distributed. This is because of linear operations, such as the first-order difference and the partial DFT, performed on the $f_\lambda^q(n)$ to determine $\mathbf{\hat{F}}_\lambda^q$. However, the variances of elements of $\mathbf{w}$ are proportional to $\Delta_B^2$ where $B$ denotes the number of quantization bits and $\Delta_B$ is the quantization step size. Note that any estimation of sparse vector $\mathbf{\hat{z}}$ from methods, such as LASSO, would result in an error bounded by $\Delta_B^2$. Specifically, we have that
\begin{align}
    \|\mathbf{\hat{z}} -  \mathbf{\hat{z}}_{_{\text{LASSSO}}} \|_2^2 \leq C_{_{\text{LASSSO}}} \, \Delta_B^2,
\end{align}
where $\mathbf{\hat{z}}_{_{\text{LASSSO}}}$ is an estimate of $\mathbf{\hat{z}}$ and $C_{_{\text{LASSSO}}}$ is a constant \cite[Ch. 1]{LIP2}.

Next, consider a scenario where the support of the sparse vector is known. In this case, the LASSO problem boils down to a least-squares regression. In particular, let $\mathcal{T} = \text{supp}(\hat{\mathbf{z}})$, $\mathcal{T} \subset \{1,2,\dots,N\}$, and $L = \#\mathcal{T} \ll M < N$. With the support information, it only requires estimating the amplitudes of the non-zero values of $\mathbf{\hat{z}}$. This is estimated as 
\begin{align}
     \mathbf{\hat{z}}_\mathcal{T}^e = \mathbf{V}_\mathcal{T}^\dagger\mathbf{\hat{F}}_\lambda ^q = \left(\mathbf{V}_\mathcal{T}^H \mathbf{V}_\mathcal{T}\right)^{-1}\mathbf{V}_\mathcal{T}^H \mathbf{\hat{F}}_\lambda^q.
\end{align}
The estimation is referred to as \emph{oracle estimator} as it requires the knowledge of the support. The estimator is computationally simpler than the LASSO estimation. In addition, it requires fewer measurements than LASSO, as support is already known.

The support of $\mathbf{\hat{z}}$ indicates the time locations around which the folding of $f_\lambda(t)$ happens. Suppose this information is provided by the modulo-ADC hardware; then, it requires one bit to encode it. Given that the ADC has $B$ bits, and one bit (Least Significant Bit-LSB) is reserved to indicate the support, only $B-1$ bits can be used to quantize $f_\lambda(n)$. In this case, the variance of the quantization noise would be four times that in the $B$-bit quantizer. Hence, if $\mathbf{\hat{z}}_{_{\text{LS}}}$ estimate of $\mathbf{\hat{z}}$, determined from the support $\mathcal{T}$ and $\mathbf{\hat{z}}_\mathcal{T}^e$, then it can be shown that
\begin{align}
    \|\mathbf{\hat{z}} -  \mathbf{\hat{z}}_{_{\text{LS}}} \|_2^2 \leq C_{_{\text{LS}}} \, 4\Delta_B^2,
\end{align}
where $C_{_{\text{LS}}}$ is a constant term \cite[Ch. 1]{LIP2}. Hence, an LS approach is feasible by allocating one bit for support and the rest for quantizing the samples. 
In the following, the bit-distribution method is compared with the existing methods through simulations, and we show that it performs favorably. Then we discuss the hardware implementation of the proposed approach.

\subsection{Simulations}
\begin{figure*}[t!]
    \centering
    \includegraphics[height = 7.5cm, width = 18cm]{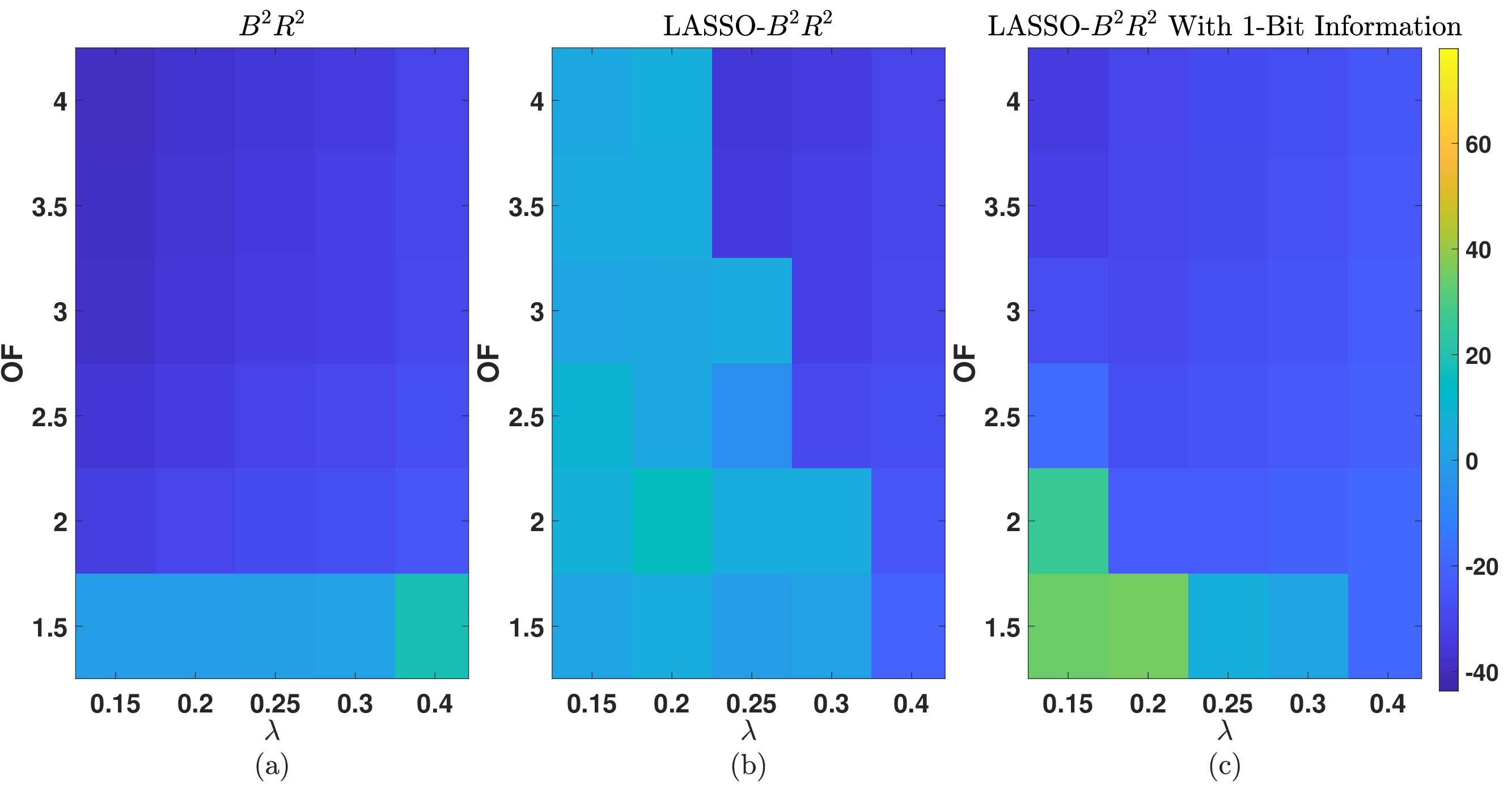}
    \caption{(a)-(c) Average NMSE computed for various $\lambda$ and OF pairs using B$^2$R$^2$, LASSO-B$^2$R$^2$, and LASSO-B$^2$R$^2$ with 1-bit, respectively.}
    \label{fig:Comparison2}
\end{figure*}
\begin{figure*}[t!]
    \centering
    \includegraphics[height = 9.5cm, width = 16cm]{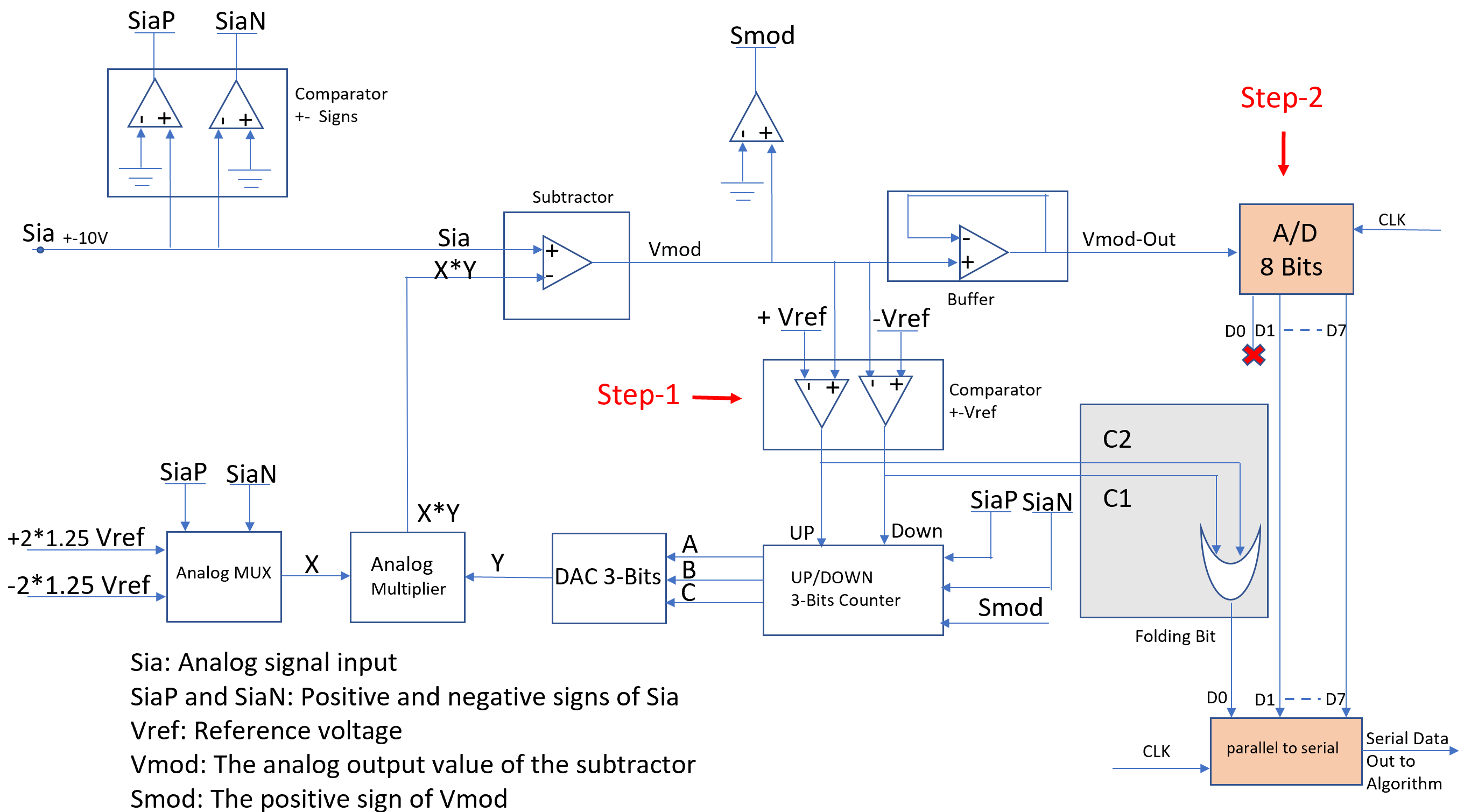}
    \caption{Hardware Prototype of modulo-ADC with additional 1-bit information.}
    \label{fig:HW_Prototype}
\end{figure*}
In this section, we evaluate the performance of LASSO-B$^2$R$^2$ with 1-bit and compare it with B$^2$R$^2$ and LASSO-B$^2$R$^2$. The analysis is based on the BL signal shown in Fig. \ref{fig:Comparison}(a), which has a maximum frequency component of $f_m = 1000$ Hz. This signal is passed through a modulo operator with a threshold $\lambda = 0.2$. Fig. \ref{fig:Comparison}(b) illustrates the sampled and quantized modulo signal, with an OF$=3$ and quantization using $b=6$ bits. In comparison, Fig. \ref{fig:Comparison}(c) shows the quantized modulo signal for $b=5$ bits, along with an additional 1-bit indicator that is set to one at each level-crossing and zero otherwise. We aim to recover the true unfolded signal from these quantized modulo samples using B$^2$R$^2$, LASSO-B$^2$R$^2$, and LASSO-B$^2$R$^2$ with 1-bit information, with the reconstruction results shown in Fig. \ref{fig:Comparison}(d)-(f), respectively. Table \ref{Tab:2} presents the MSE and the computational time required required by each algorithm for this example.
Based on these results, we observe that LASSO-B$^2$R$^2$ with 1-bit information successfully recovers the signal for lower $\lambda$ values, where the LASSO-B$^2$R$^2$ algorithm fails. Notably, while B$^2$R$^2$ also recovers the true signal, it requires significantly more computational time. 
\begin{table}[!ht]
    \centering
    \caption{\small{Comparison of B$^2$R$^2$, LASSO-B$^2$R$^2$, and LASSO-B$^2$R$^2$ with 1-bit in terms of MSE and computational complexity.}}
    \begin{tabular}{|M{1.3cm}|M{1.2cm}|M{1.8cm}|M{2cm}|}
    \hline
        ~ & B$^2$R$^2$ & LASSO-B$^2$R$^2$ & LASSO-B$^2$R$^2$ With 1-Bit \\ \hline
        MSE (dB) & -12.8431 & 9.7573 & -9.0683\\ \hline
        Time (Sec) & 5.5757 & 2.1289 &  0.0029 \\ \hline
    \end{tabular}
    \label{Tab:2}
\end{table}

To further validate the proposed algorithms, we evaluate their performance across various combinations of $\lambda$ and OF. For each ($\lambda$, OF) pair, 100 signal realizations are generated, and the proposed algorithms are applied. Figures \ref{fig:Comparison2}(a)-(c) illustrate the average NMSE results obtained for B$^2$R$^2$, LASSO-B$^2$R$^2$, LASSO-B$^2$R$^2$ with 1-bit, respectively.
The results indicate that LASSO-B$^2$R$^2$ with 1-bit addresses the performance limitations of LASSO-B$^2$R$^2$ at lower $\lambda$ values, while maintaining a performance level comparable to B$^2$R$^2$. Moreover, as shown in Table \ref{tab:Comparison1}, LASSO-B$^2$R$^2$ with 1-bit achieves this with significantly lower computational time compared to B$^2$R$^2$.
\begin{table}[h!]
\caption{Average computational time (in seconds) required by each algorithm for various $\lambda$ and OF pairs.}
\centering
\begin{tabular}{|c|c|c|c|c|c|}
\hline
\multicolumn{1}{|c|}{} & \multicolumn{5}{c|}{B$^2$R$^2$} \\ \hline
\textbf{OF\textbackslash $\lambda$} & 
0.15 & 0.2 & 0.25 & 0.3 & 0.4 \\ \hline
1.5 & 11.7611 & 8.3891 & 8.2047 & 8.3537 & 8.3337 \\ \hline
2 & 1.3693 & 1.1471 & 0.9964 & 0.7903 & 0.6315 \\ \hline
2.5 & 1.1012 & 0.8338 & 0.6785 & 0.7367 & 0.6439  \\ \hline
3 & 0.9505 & 0.7044 & 0.7532 & 0.6912 & 0.6719 \\ \hline
3.5 & 0.7887 & 0.7411 & 0.7152 & 0.7322 & 0.5924  \\ \hline
4 & 0.8017 & 0.8083 & 0.7506 & 0.6175 & 0.5566  \\ \hline

\multicolumn{1}{|c|}{\textbf{}} & \multicolumn{5}{c|}{LASSO-B$^2$R$^2$} \\ \hline
\textbf{OF\textbackslash $\lambda$} & 0.15 & 0.2 & 0.25 & 0.3 & 0.4 \\ \hline
1.5 & 0.5449 & 0.3436 & 0.2724 & 0.2765 & 0.6130  \\ \hline
2 & 0.2002 & 0.9797 & 0.1294 & 0.1033 & 0.0946 \\ \hline
2.5 & 1.0606 & 0.4519 & 0.4286 & 1.3876 & 0.0773  \\ \hline
3 & 0.8238 & 0.2158 & 1.0611 & 0.1305 & 0.0836\\ \hline
3.5 & 0.1341 & 0.1278 & 0.1232 & 0.0695 & 0.0406 \\ \hline
4 & 0.1354 & 0.2128 & 0.0714 & 0.0506 & 0.0340 \\ \hline

\multicolumn{1}{|c|}{\textbf{}} & \multicolumn{5}{c|}{LASSO-B$^2$R$^2$ With 1-Bit Information} \\ \hline
\textbf{OF\textbackslash $\lambda$}& 0.15 & 0.2 & 0.25 & 0.3 & 0.4 \\ \hline
1.5 & 0.0021 & 0.0003 & 0.0004 & 0.0003 & 0.0002\\ \hline
2 & 0.0004 & 0.0005 & 0.0004 & 0.0001 & 0.0001  \\ \hline
2.5 & 0.0021 & 0.0016 & 0.0006 & 0.0012 & 0.0009\\ \hline
3 & 0.0020 & 0.0005 & 0.0009 & 0.0008 & 0.0006 \\ \hline
3.5 & 0.0007 & 0.0005 & 0.0007 & 0.0005 & 0.0001 \\ \hline
4 & 0.0013 & 0.0004 & 0.0005 & 0.0005 & 0.0003 \\ \hline
\end{tabular}
\label{tab:Comparison1}
\end{table}

\subsection{Hardware Prototype}
\begin{figure*}[t!]
    \centering
    \includegraphics[height = 12cm, width = 18cm]{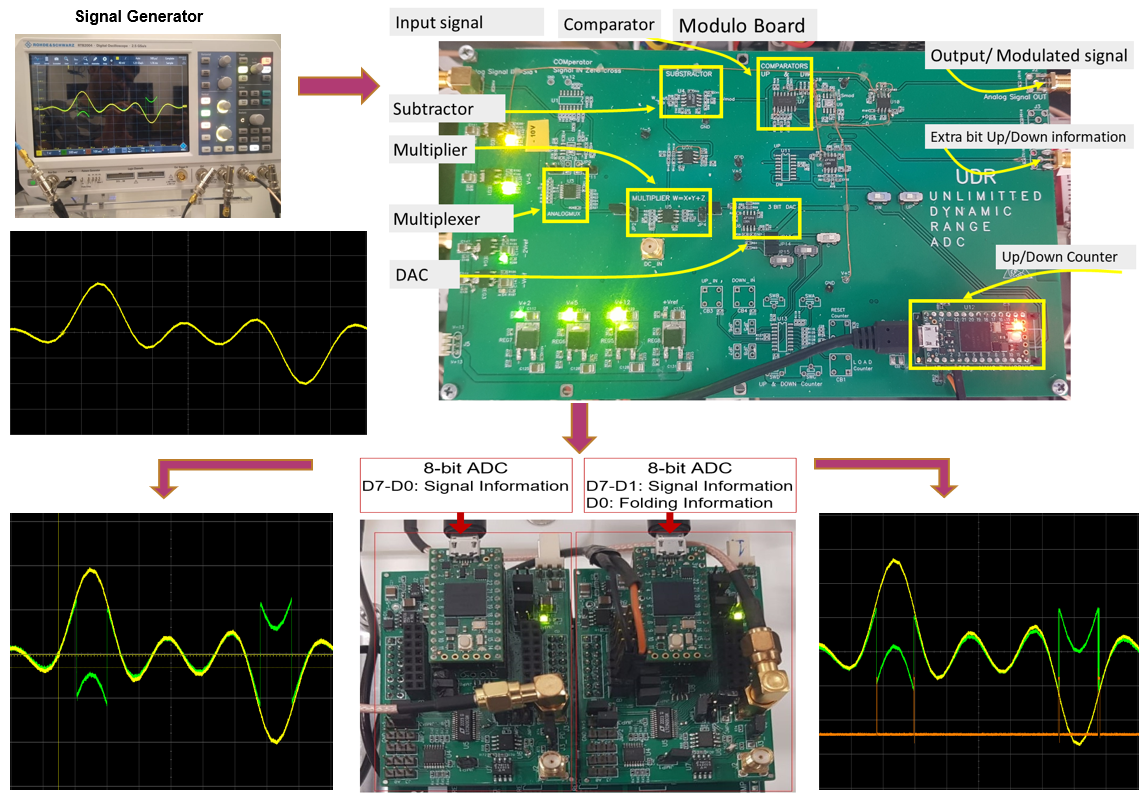}
    \caption{\small{Hardware Setup: A Modulo Board is paired with two 8-bit ADCs. One ADC utilizes all 8 bits to capture the signal information, while the other ADC uses its LSB to record the folding instance or level-crossing and the remaining bits to preserve the signal information.}}
    \label{fig:HW}
\end{figure*}
In this section, we present the hardware implementation of the proposed LASSO-B$^2$R$^2$ with 1-bit information. The hardware is built upon the modulo board developed in \cite{Mod_HW}. In particular, the gray-shaded region in Fig. \ref{fig:HW_Prototype} highlights the additional circuitry (an OR gate) introduced to augment the original setup in \cite{Mod_HW} for capturing the 1-bit information. For a complete overview of the initial modulo hardware prototype, please refer to Section 3 of \cite{Mod_HW}. The working principle of the added circuitry is described as follows:
\begin{itemize}
    \item \textbf{\textcolor{red}{Step-1:}} The intermediate analog modulo signal, Vmod, is compared against thresholds $+\text{Vref}=+\lambda$ and $-\text{Vref}=-\lambda$ using two comparators. Each comparator outputs a binary value (0 or 1) based on the signal level:
    \begin{itemize}
        \item If $\text{Vmod} > +\text{Vref}$, the first comparator’s output, $C_1$, is $0$, and the second comparator’s output, $C_2$, is $1$.
        \item If $\text{Vmod} < -\text{Vref}$, then $C_1 = 1$ and $C_2=0$.
        \item If $\text{Vref} > \text{Vmod} > -\text{Vref}$, then $C_1 = 0$ and $C_2 = 0$.
    \end{itemize}
     Here, a value of 1 in either $C_1$ or $C_2$ indicates a folding operation has occurred.
    \item \textbf{\textcolor{red}{Step-2:}} An 8-bit ADC then converts the analog modulo signal, Vmod-Out, to digital bits. After digitizing each sample, we discard the LSB, denoted D0, and replace it with a folding bit obtained by an OR-gate operation between $C_1$ and $C_2$. This processed digital output is then serialized using a parallel-to-serial converter and fed into the recovery algorithm.
\end{itemize}
The modulo hardware board along with two 8-bit ADCs is depicted in Fig. \ref{fig:HW}. 
Here, we generate a $1$K Hz BL signal using the waveform generator in a digital storage oscilloscope. The output from the modulo board is then sampled (with OF=4) and quantized by two 8-bit ADCs. One ADC dedicates all its 8 bits to capture the signal information, while the other ADC reserves the LSB to store the folding instant or level-crossing information, with the remaining bits capturing the signal information.
\begin{table}[!ht]
    \centering
    \caption{Hardware results}
    \begin{tabular}{|M{1.3cm}|M{1.2cm}|M{1.8cm}|M{2cm}|}\hline
        ~ & B$^2$R$^2$ & LASSO-B$^2$R$^2$ & LASSO-B$^2$R$^2$ With 1-Bit \\ \hline
        MSE (dB) & 7.8149 & 7.8149 & 4.8 \\ \hline
        Time (Sec) & 0.2938 & 0.0363 &  0.00074 \\ \hline
    \end{tabular}
    \label{Tab:3}
\end{table}
We now aim to recover the true unfolded signal from these quantized modulo samples using the above mentioned algorithms. Table \ref{Tab:2} tabulates the MSE and the computational time required by each algorithm.
The results demonstrate that LASSO-B$^2$R$^2$ with 1-bit information successfully recovers the signal with a substantially lower computational time than both B$^2$R$^2$ and LASSO-B$^2$R$^2$.

Based on the \textit{bits distribution} mechanism introduced in this work, the authors in \cite{Bernardo} analyzed a modulo-ADC within a dithered quantization framework, deriving a closed-form expression for the MSE between the true and estimated samples. They also compared its performance with that of a standard ADC. 


 

\section{Conclusion}
In this paper, we proposed recovery algorithms that operate near the Nyquist rate, designed to enhance accuracy, computational efficiency, and robustness in high dynamic range (DR) signal acquisition for modulo-ADC systems. Building upon existing modulo recovery algorithms, we introduced LASSO-B$^2$R$^2$, a sparse residual recovery algorithm that leverages the sparsity of the first-order difference of residual samples. Numerical simulations demonstrated that LASSO-B$^2$R$^2$ outperformed existing methods in terms of speed and robustness. However, at lower modulo thresholds, it required slightly higher sampling rates. To overcome this limitation, we introduced the \textit{bits distribution} mechanism, which assigned 1-bit to identify modulo folding events. This mechanism simplifies the recovery problem to a pseudo-inverse computation, significantly improving computational efficiency. Numerical simulations validated the superiority of our approach in terms of computational efficiency and reconstruction accuracy. Additionally, we developed a hardware prototype to acquire the 1-bit folding information, demonstrating the practicality of the proposed algorithm.

\section*{Acknowledgement}
The authors gratefully acknowledge Shlomi Savariego, Moshe Namer, and Nimrod Glazer for their invaluable assistance and support in developing the hardware prototype.

\ifCLASSOPTIONcaptionsoff
  \newpage
\fi
\bibliographystyle{ieeetr}
\bibliography{bibs_1,US_biblios}
\end{document}